\documentclass{emulateapj_modified}
\def\deg{\ifmmode^\circ\else$^\circ$\fi}

\def\lsun{L$_{\odot}$}

\def\arcs{\ifmmode {''}\else $''$\fi}
\def\arcm{\ifmmode {'}\else $'$\fi}
\def\parcs{\sa=.07em \sb=.03em
     \ifmmode $\rlap{.}$^{\scriptscriptstyle\prime\kern -\sb\prime}$\kern -\sa$
     \else \rlap{.}$^{\scriptscriptstyle\prime\kern -\sb\prime}$\kern -\sa\fi}
\def\parcm{\sa=.08em \sb=.03em
     \ifmmode $\rlap{.}\kern\sa$^{\scriptscriptstyle\prime}$\kern-\sb$
     \else \rlap{.}\kern\sa$^{\scriptscriptstyle\prime}$\kern-\sb\fi}

\def\spose#1{\hbox to 0pt{#1\hss}}
\def\simlt{\mathrel{\spose{\lower 3pt\hbox{$\mathchar"218$}}
     \raise 2.0pt\hbox{$\mathchar"13C$}}}
\def\simgt{\mathrel{\spose{\lower 3pt\hbox{$\mathchar"218$}}
     \raise 2.0pt\hbox{$\mathchar"13E$}}}
\def\lsim{\rlap{$<$}{\lower 1.0ex\hbox{$\sim$}}}
\def\gsim{\rlap{$>$}{\lower 1.0ex\hbox{$\sim$}}}

\begin{document}

\title{GOALS: The Great Observatories All-sky LIRG Survey}

\author{L. Armus\altaffilmark{1}, J.M. Mazzarella\altaffilmark{2}, A.S. Evans\altaffilmark{3,4}, J.A. Surace\altaffilmark{1}, D.B. Sanders\altaffilmark{5}, K. Iwasawa\altaffilmark{6}, D.T. Frayer\altaffilmark{7}, J.H. Howell\altaffilmark{1}, B. Chan\altaffilmark{2}, A.O. Petric\altaffilmark{1}, T. Vavilkin\altaffilmark{8}, D.C. Kim\altaffilmark{3}, S. Haan\altaffilmark{1}, H. Inami\altaffilmark{9}, E.J. Murphy\altaffilmark{1}, P.N. Appleton\altaffilmark{6}, J.E. Barnes\altaffilmark{4}, G. Bothun\altaffilmark{10}, C.R. Bridge\altaffilmark{1}, V. Charmandaris\altaffilmark{11}, J.B. Jensen\altaffilmark{12}, L.J. Kewley\altaffilmark{4}, S. Lord\altaffilmark{2}, B.F. Madore\altaffilmark{2,13}, J.A. Marshall\altaffilmark{14}, J.E. Melbourne\altaffilmark{15}, J. Rich\altaffilmark{4}, S. Satyapal\altaffilmark{16}, B. Schulz\altaffilmark{6}, H.W.W. Spoon\altaffilmark{17}, E. Sturm\altaffilmark{18}, V. U\altaffilmark{4}, S. Veilleux\altaffilmark{19}, K. Xu\altaffilmark{6}}

%\altaffiltext{1}{based on observations obtained with the Spitzer Space Telescope, which is
%operated by the Jet Propulsion Laboratory, California Institute of Technology, under NASA 
%contract 1407}
\altaffiltext{1}{Spitzer Science Center, Calfornia Institute of Technology, MS 220-6, Pasadena, CA 91125}
\altaffiltext{2}{Infrared Processing and Analysis Center, Calfornia Institute of Technology, MS 100-22, Pasadena, CA 91125}
\altaffiltext{3}{National Radio Astronomy Observatory, 520 Edgemont Road, Charlottesville, VA 22903}
\altaffiltext{4}{Department of Astronomy, University of Virginia, P.O. Box 400325, Charlottesville, VA 22904}
\altaffiltext{5}{Institute for Astronomy, University of Hawaii, 2680 Woodlawn Drive, Honolulu, HI 96822}
\altaffiltext{6}{INAF-Observatorio Astronomico di Bologna, Via Ranzani 1, Bologna, Italy}
\altaffiltext{7}{NASA Herschel Science Center, California Institute of Technology, MS 100-22, Pasadena, CA 91125}
\altaffiltext{8}{Department of Physics and Astronomy, SUNY Stony Brook, Stony Brook, NY, 11794}
%\altaffiltext{8}{Institute of Space and Astronautical Science, Japan Aerospace Exploration Agency, Japan}
\altaffiltext{9}{Department of Space and Astronautical Science, The Graduate University for Advanced Studies, Japan}
\altaffiltext{10}{Physics Department, University of Oregon, Eugene OR, 97402}
\altaffiltext{11}{Department of Physics, University of Crete, P.O. Box 2208, GR-71003, Heraklion, Greece}
\altaffiltext{12}{Gemini Observatory, 940 N. Cherry Ave., Tucson, AZ, 85719}
\altaffiltext{13}{The Observatories, Carnegie Institute of Washington, 813 Santa Barbara Street, Pasadena, CA 91101}
\altaffiltext{14}{The Jet Propulsion Laboratory, California Institute of Technology, Pasadena, CA 91125}
\altaffiltext{15}{California Institute of Technology, MS 320-47, Pasadena, CA 91125}
\altaffiltext{16}{Department of Physics \& Astronomy, George Mason University, 4400 University Drive, Fairfax, VA 22030}
\altaffiltext{17}{Department of Astronomy, Cornell University, Ithaca, NY, 14953}
\altaffiltext{18}{MPE, Postfach 1312, 85741 Garching, Germany}
\altaffiltext{19}{Astronomy Department, University of Maryland, College Park, MD 20742}
\altaffiltext{*}{http://goals.ipac.caltech.edu}

%\altaffiltext{8}{The IRS was a collaborative venture between Cornell University and 
%Ball Aerospace Corporation funded by NASA through the Jet Propulsion Laboratory and
%the Ames Research Center}

\begin{abstract}

The Great Observatories All-sky LIRG Survey (GOALS$^{*}$) combines data from NASA's
Spitzer, Chandra, Hubble and GALEX observatories, together with ground-based data into a comprehensive imaging and
spectroscopic survey of over 200 low redshift ($z < 0.088$), Luminous Infrared
Galaxies (LIRGs).  The LIRGs are a complete subset of the IRAS Revised Bright
Galaxy Sample (RBGS), which comprises 629 extragalactic objects with $60\mu$m
flux densities above 5.24 Jy, and Galactic latitudes above five degrees.  The
LIRGs targeted in GOALS span the full range of nuclear spectral types defined
via traditional optical line-ratio diagrams (type-1 and type-2 AGN, LINERs, and
starbursts) as well as interaction stages (major mergers, minor mergers, and
isolated galaxies).  They provide an unbiased picture of the processes
responsible for enhanced infrared emission in galaxies in the local Universe.
As an example of the analytic power of the multi-wavelength GOALS dataset, we
present Spitzer, Chandra, Hubble and GALEX images and spectra for the interacting
system VV 340 (IRAS F14547+2449).  The Spitzer MIPS imaging data indicates that
between $80-95$\% of the total far-infrared emission (or about
$5\times10^{11}$L$_{\odot}$) originates in VV 340 North.  While the Spitzer IRAC colors
of VV 340 North and South are consistent with star-forming galaxies, both the
Spitzer IRS and Chandra ACIS data indicate the presence of an AGN in VV 340
North.  The observed line fluxes, without correction for extinction, imply the
AGN accounts for less than $10-20$\% of the observed infrared emission.  The
X-ray data are consistent with a heavily absorbed (N$_{H} \ge 10^{24}$cm$^{-2}$)
AGN.  The GALEX far and near-UV fluxes imply a extremely large infrared
``excess" (IRX) for the system (F$_{IR}$/F$_{fuv} \sim 81$) which is well above
the correlation seen in starburst galaxies.  Most of this excess is driven by
VV 340 N, which has an IR excess of nearly 400.  The VV 340 system seems to be
comprised of two very different
galaxies -- an infrared luminous edge-on galaxy (VV 340 North) 
that dominates the long-wavelength emission from the system and which 
hosts a buried AGN, and a face-on
starburst (VV 340 South) that dominates the short-wavelength emission.

\end{abstract}


\section{Introduction}

The Infrared Astronomical Satellite (IRAS) provided the first unbiased
survey of the sky at mid and far-infrared wavelengths, giving us a
comprehensive census of the infrared emission properties of galaxies in the
local Universe. The high luminosity tail of the infrared luminosity
function can be approximated by a power-law (L$_{IR} ^{-2.35}$), which
implies a space density for the most luminous infrared sources which is
well in excess of what is found in the optical for local galaxies (e.g.,
Schechter 1976). At the highest luminosities, Ultraluminous Infrared
Galaxies, or ULIRGs (those galaxies with L$_{IR} > 10^{12}$ \lsun) have a space density that is a 
factor of $1.5-2$ higher
than that of optically selected QSOs, the only other known objects
with comparable bolometric luminosities (Schmidt \& Green 1983).

Multi-wavelength imaging surveys have shown that nearly
all ULIRGs are found in systems which have undergone
strong tidal perturbations due to the merger of pairs of 
gas-rich disk galaxies (Armus et al. 1987; Sanders et al. 1988a,b; 
Murphy et al. 1996,2001a). They have enhanced star formation rates
compared to non-interacting galaxies, and the fraction of sources with
active galactic nuclei (AGN) increases as a function of increasing $L_{IR}$ 
(Armus, et al. 1989; Murphy et al. 1999,2001b; Veilleux et al. 1995,1997; Kim et al. 1998). ULIRGs may
represent an important evolutionary stage in the formation of QSOs (e.g., Sanders
et al. 1988a,b) and perhaps powerful radio galaxies (Mazzarella et al.
1993; Evans et al. 2005). In fact, numerical simulations have shown that
tidal dissipation in mergers can be very effective in driving material from
a gas-rich galaxy disk towards the nucleus, fueling a starburst and/or a
nascent AGN (e.g., Barnes \& Hernquist 1992). Morphological and kinematic
studies of ULIRGs also suggest that their stellar
populations are evolving into relaxed, elliptical-like distributions
(Wright, et al. 1990; Genzel et al. 2001; Tacconi et al. 2002).

Luminous Infrared Galaxies, or LIRGs (those galaxies with L$_{IR} \ge 10^{11}$ \lsun),
are interesting phenomena in their own right, but they also may play
a central role in our understanding of the general evolution of
galaxies and black holes, as demonstrated by two key observational
results.  First, observations with ISO and Spitzer have shown that
LIRGs comprise a significant fraction ($\ge 50\%$)of the cosmic infrared background and dominate
the star-formation activity at $z\sim 1$ (Elbaz et al. 2002; Le Floc'h
et al. 2005; Caputi et al. 2007, Magnelli et al. 2009).  
In comparison to the local universe where they are relatively
rare, ULIRGs are about a thousand times more common at $z\ga2$
(Blain et
al. 2002; Chapman et al. 2005).  Second, the ubiquity of super-massive
nuclear black holes in quiescent galaxies and the scaling of their
masses with stellar bulge masses (Magorrian et al. 1998; Ferrarese \& Merritt 2000; 
Gebhardt et al. 2000) suggest
an intimate connection between the evolution of massive galaxies and
their central black holes.  One explanation for the relationship is
that mass accretion onto the black hole occurs during episodes of
nuclear star formation.  The study of a large, complete sample of local
LIRGs spanning all merger and interaction stages can shed critical light on
the co-evolution of black holes and stellar bulges in
massive galaxies.

The Great Observatories All-sky LIRG Survey (GOALS) combines data from NASA's
Spitzer, Hubble, Chandra and GALEX observatories in a comprehensive imaging
and spectroscopic survey of over 200 low redshift ($z < 0.088$) LIRGs (see Table 1).
The primary new data sets consist of Spitzer IRAC and
MIPS imaging, IRS spectroscopy, HST ACS, NICMOS, and WFPC2 imaging,
Chandra ACIS imaging, and GALEX far and near-UV observations.  The majority 
of the new observations are being led by GOALS team members, but 
we also make use of the Spitzer, HST, Chandra and GALEX archives to fill out the 
sample data (see Table 2).  In addition, optical and K-band imaging (Ishida et al. 2004), 
optical spectra (Kim et al. 1995), J, H, and Ks-band near-infrared images from 2MASS 
(Skrutskie et al. 2006),  sub-millimeter images (Dunne et al. 2000), CO and HCN
(Sanders, Scoville \& Soifer 1991; Gao \& Solomon 2004), and 20cm VLA imaging (Condon et al. 1990) exist for
various sub-sets of the LIRGs in GOALS, besides the IRAS data with which the 
sources were selected (Sanders et al. 2003).
In this paper we introduce the main components of the survey, present our primary 
science objectives, and discuss the multi-wavelength
results for one source, VV 340.  The layout of the paper is as follows.
The scientific objectives are discussed in 
section 2, followed by a definition of the sample in section 3.  In 
section 4 we describe the observations, in section 5 we outline the Spitzer
Legacy data products being delivered, and in section 6 we present results
for the LIRG VV 340.

\section{Scientific Objectives of GOALS}

While a great deal of effort has been devoted to the study of ULIRGs, LIRGs
have, in comparison, suffered from a lack of attention.  Ground-based optical
and near infrared imaging studies of low-redshift LIRGs have been performed
(e.g., Ishida 2004), but multi-wavelength imaging and spectroscopic surveys of a
large sample of the nearest and brightest LIRGs have not been completed.  This
is undoubtedly due to the fact that LIRGs form a morphologically diverse group
of galaxies, unlike ULIRGs which are nearly always involved in the final stages
of a violent and spectacular merger.  LIRGs are also much harder to detect and
analyze at high redshift in many shallow surveys, which has traditionally
removed much of the impetus behind establishing a robust library of low-redshift
analogs.  This has changed recently, with surveys such as COSMOS (Scoville et
al. 2007) and GOODS (Dickinson et al. 2003),
which include significant samples of LIRGs at
cosmologically interesting epochs.

An important question regarding the nature of LIRGs, is what is the source of
their power?  What mechanism is responsible for generating energy at a rate that
is tens to hundreds of times larger than emitted by a typical galaxy?
Interactions between large, late-type galaxies are evident in many systems, but
a significant fraction of LIRGs show no morphological evidence (e.g., double
nuclei, tidal tails or stellar bridges, etc.) for a major merger.  Large
molecular gas masses, large gas fractions, or unusual dynamics may all play a
role.  Regardless of the trigger, an AGN or an intense starburst is often
invoked as the most likely source of the energy in LIRGs.  The relative
importance of AGN and starburst activity within individual systems, for the
generation of far-infrared radiation from LIRGs as a class, 
and how the dominant energy source changes as a function of merger stage are all important
questions that can only be answered by studying a large, unbiased sample at a
wide variety of wavelengths.  
This is the primary motivation behind the GOALS
project.

By their very nature, LIRGs are dusty galaxies, wherein a large fraction (over
90\% for the most luminous systems) of the UV light emitted by stars and/or AGN
is absorbed by grains and re-radiated in the far-infrared.  This makes
traditonal, UV and optical spectroscopic techniques for detecting and measuring
starbursts and AGN (e.g., emission line ratios and linewidths) difficult to link
quantitatively to the ultimate source of the IR emission.  However, by combining
these data with observations in the infrared, hard x-ray, and radio, it is
possible to penetrate the dust and better understand the LIRG engine.

With the IRAS data we have been able to construct unbiased, flux-limited samples
of LIRGs and measure their global far-infrared colors, albeit with little or no
information about the distribution of the dust, due to the large IRAS
beam sizes.  The overall LIRG Spectral Energy Distributions (SEDs) are shaped by the
relative locations of the dust and the young stars, as well as the presence of
an active nucleus.  At a basic level, the size of the warm dust emiting region
can give an indication of the power source, since a starburst will be spread out
on kpc scales while the region heated by an AGN will be $2-3$ orders of
magnitude smaller.  Spitzer imaging with IRAC and MIPS can give us our first
estimates of the size of the IR-emitting regions in the nearest LIRGs, although
higher spatial resolution is required for the vast majority of systems if we are
to place meaningful constraints on, for example, the luminosity density of the
nuclear starburst.  Ground-based $10-20\mu$m imaging can provide a high spatial
resolution view of the nucleus, but these wavelengths are most sensitive to the
hottest dust, and extrapolations to the far-infrared are always required.  The
IRAC data provides the finest spatial resolution on Spitzer, and since the IRAC
bands for low-redshift galaxies are dominated by stars, hot dust, and finally
PAH emission as one progresses from $3.6\mu$m to $8\mu$m, the nuclear colors
themselves can indicate the presence of an AGN (Lacy et al. 2004; Stern et al.
2005).  Although the spatial resolution of the MIPS data is limited, the flux
ratios give an indication of the dust temperature and hence, the importance of
an AGN to the overall energy budget.  When coupled with the radio continuum
data, the MIPS fluxes can also be used to estimate the radio--far-infrared flux
ratio, itself an indicator of the presence of a (radio loud) AGN.

Mid-infrared and X-ray spectroscopy provide powerful tools for not only
uncovering buried AGN, but also measuring the physical properties of the gas and
dust surrounding the central source.  The low-resolution IRS spectra are ideal
for measuring the broad PAH features at 6.2, 7.7, 8.6, 11.3, and $12.7\mu$m, the
silicate absorption features at 9.7 and $18\mu$m, and water or hydrocarbon ices
at $5-7\mu$m (e.g., Spoon et al. 2004; Armus et al. 2007).  The PAH ratios
indicate the size and ionization state of the small grains (e.g., Draine \& Li
2001), while the silicate absorption gives a direct measure of the optical depth
towards the nuclei and clues as to the geometry of the obscuring medium (e.g.
Levenson et al. 2007, Sirocky et al. 2008).  The ratio of the PAH emission to
the underlying (hot dust) continuum can be used as a measure of the strength of
an AGN, since Seyert galaxies and quasars typically have extremely low PAH
equivalent widths (EQW) compared to starburst galaxies (Genzel et al. 1998;
Sturm et al. 2002, Armus et al. 2004,2006,2007).  Among ULIRGs, the PAH EQW have
even been shown to decrease with increasing luminosity (Tran et al. 2001; Desai
et al. 2007) suggesting an increasing importance of an AGN in the highest
luminosity systems.  Since the strengths of the PAH features can dominate
broad-band, mid-infrared filter photometry, it is critical to understand how the
relative strengths of the PAH features vary within the LIRG population and how
they affect the colors with redshift (see Armus et al.  2007).  In addition, the
shape of the silicate absorption can indicate the presence of crystalline
silicates, themselves a clear sign of violent, and recent, grain processing
(e.g., Spoon et al. 2006).

The high-resolution IRS spectra are ideally suited for measuring the equivalent
widths and relative line strengths of the narrow, fine-structure atomic emission
lines such as [S IV] $10.5\mu$m, [Ne II] $12.8\mu$m, [Ne III] $15.5\mu$m, [Ne V]
14.3 and $24.3\mu$m , [O IV] $25.9\mu$m, [S III] 18.7 and $33.5\mu$m , [Si II]
$34.8\mu$m, [Fe II] 17.9 and $25.9\mu$m, as well as the pure rotational H$_2$
lines at $9.66\mu$m, $12.1\mu$m, $17.0\mu$m and $28.2\mu$m.  These lines provide
a sensitive measure of the ionization state and density of the gas, and they can
be also used to infer the basic properties (e.g., T$_{eff}$) of the young
stellar population.  The H$_2$ lines probe the warm (100-500 K) molecular gas
and, when combined with measurements of the cold molecular gas via the mm CO
line, can be used to infer the warm-to-cold molecular gas fraction as a function
of evolutionary state (e.g. Higdon et al. 2007).  The high resolution spectra
can also indicate the presence of a warm, interstellar medium, via absorption
features of C$_2$H$_2$ and HCN at $13.7\mu$m and $14.0\mu$m, respectively (see
Lahuis et al. 2007).

The low background of the ACIS detector, together with the excellent spatial
resolution of Chandra, provide an unprecedented view of the $0.4-8$ keV X-ray
emission in LIRGs.  X-ray binaries, an AGN, and hot gas associated with an
outflowing wind can all contribute to the measured X-ray emission on different
physical scales and at different characteristic energies.
The X-ray images will allow us to quantify the amount and extent of
resolved X-ray emission in LIRGs as an indication of extent of the starburst,
and/or the presence of an outflowing wind. The winds are driven through the
combined action of overlapping supernovae, producing both a hot central
component and soft, extended emission, made up predominantly of shocked,
swept-up gas in the ISM (e.g., Fabbiano, Heckman \& Keel 1990; Heckman et al.
1990; Armus et al. 1995; Read, Ponman \& Strickland 1997; Dahlem, Weaver \&
Heckman 1998; Strickland et al. 2004a,b).  An AGN, on the other hand, will produce unresolved hard
X-ray emission, sometimes accompanied by a strong neutral or ionized iron line.

The penetrating power of the hard X-rays implies that even obscured AGN should
be visible to Chandra (e.g., Komossa et al. 2003).  Unresolved, hard X-ray
($2-8$ keV) flux with strong iron line emission is a clear indication of an AGN,
and by fitting the spectrum we can derive both the intrinsic luminosity and the
HI column toward the accretion disk(s).  Of course, identifying the true nature
of the nuclear emission from the Chandra spectra alone is difficult for sources
where the HI column density exceeds $\sim 10^{24}$cm$^{-2}$, but the combination
of X-ray spectral and spatial information can help disentangle the extended
starburst emission from the unresolved, harder emission from an AGN in many
LIRGs.  Of particular interest among the LIRG sample is whether or not the X-ray
luminosity can be used as a quantitative measure of the star formation rate.  In
galaxies dominated by star formation, where the X-ray emission is dominated by
high-mass X-ray binaries, there is a correlation between the X-ray luminosity
and the star formation rate as measured by the infrared or radio emission
(Ranalli et al. 2003; Grimm et al.  2003).  It will be interesting to determine
whether or not this relation holds for the powerful starbursts found in LIRGs.

The radio emission from galaxies is immune to the effects of dust
obscuration, and the far-infrared to radio flux ratio, $q$ (Helou et al. 1985),
shows a tight correlation for star-forming galaxies (Yun, Reddy \& Condon 2001).
In addition, the
radio spectral index and the radio morphology can be used to identify buried
AGN and/or jets in the nuclei of LIRGs at the high spatial resolution afforded
with the VLA.  The available GHz radio data for the GOALS sample (Condon et al.
1990) will therefore provide a critical baseline for understanding the power sources and
the ISM properties of the GOALS sample.  In particular, it will be valuable to
compare the radio and far-infrared morphologies of the nearest LIRGs and explore
the far-infrared -- radio correlation in detail within luminous infrared
galaxies, such has been done for other nearby samples (e.g., Murphy et al. 2006).

In addition to uncovering buried AGN and nuclear starbursts in LIRGs, GOALS
provides an excellent opportunity to study star-formation in LIRG disks
as a function of merger stage.  The wide wavelength coverage and sensitivity
provide a detailed look at the old and young stars, the dust, and the gas across
the merger sequence.  GOALS will allow us to answer some important questions
about galactic mergers, such as, what fraction of the star-formation occurs in
in extra-nuclear (super) star clusters?  Can the ages and locations of these
clusters be used to reconstruct the merger history and refine detailed models of
the merger process and what are the physical processes that drive star-formation
on different scales in the merger?  A simple question that still needs to be
addressed is the range in UV properties of LIRGs.  Are LIRGs as a class, weak
or strong UV emitters?  How much of the ionizing flux emerges, and is the 
ratio of IR to UV emission related to any of the other properties, e.g. the 
merger state or the distribution of the dust?

GALEX has provided our first look at the properties of large numbers of UV-selected
starburst and AGN as a function of cosmic epoch (e.g., Gil de Paz, et al. 2007; Xu, et al. 
2007; Martin, et al. 2007; Schiminovich, et al. 2007).  By combining the GALEX and 
Spitzer data, we will be able to make a full accounting of the energy balance
throughout the GOALS sample, relating the UV emission to the infrared photometric
and spectroscopic properties of a complete sample of low-redshift LIRGs.
In the GOALS
targets that are resolved by GALEX (a small but important sub-sample), we will
be able to separate the nucleus and disk in single systems, and the individual
galaxies in interacting systems, in order to measure the relative UV emission
escaping from the systems and directly compare this to the far-infrared emission
measured with Spitzer.  This will provide a detailed look at the variation in
the IR to UV ratio within interacting galaxies at all stages. 

The HST imaging data in GOALS provides our sharpest look at the stellar
clusters and detailed morphological features in LIRGS.  The depth and high
spatial resolution of the HST imaging with the Advanced Camera for Surveys (ACS)
Wide Field Camera (WFC) allows for a sensitive search for faint remnants of
past, or less-disruptive (minor) mergers (e.g., shells, fading tails, etc.),
small scale nuclear bars, and even close, double nuclei, although the latter are
notoriously difficult to measure at optical wavelengths in heavily obscured
systems.  These data will be used to fit models of galaxy surface brightness
profiles with standard programs (e.g., GALFIT) and estimate basic structural
parameters (e.g. bulge to disk ratio, half-light radius) for the LIRGs as a
class, and as a function of nuclear properties (e.g., AGN or starburst dominated
spectra as determined from the Spitzer IRS and Chandra data).

The sub-arcsec resolution of the ACS also allows a sensitive search for and
quantitative measurement of the young, star-forming clusters in LIRGs.  The
cluster populations can be placed on color magnitude diagrams in order to
estimate their ages, and the number and luminosity of such systems can be
studied as a function of the relative age of the merger.  Particularly powerful
in studying the cluster ages will be the combination of visual and UV imaging
with ACS (the latter obtained with the Solar Blind Channel, SBC) since the
colors will then allow us to break the age-reddening degeneracy over a large
range in cluster ages ($10^{6} - 10^{8}$ yrs).  The cluster results can then be
compared to models of the interaction, as has been done in some well-studied
mergers (e.g., NGC 4038/9 -- Whitmore \& Schweizer 1995; Whitmore, et al. 2005,2007; Mengel, et al. 
2001,2005).  For example, it has
recently been shown that while the interacting LIRG NGC 2623 posesses a large
number of young star clusters that probably formed in the interaction, they
account for a negligible fraction ($ < 1\%$) of the bolometric luminosity (Evans
et al. 2008).

Our understanding of the UV emission properties of LIRGs will be greatly
enhanced by the addition of the high-resolution ACS far-UV imaging data.  Only a
handful of LIRGs have been imaged in the vacuum ultraviolet at high
resolution (e.g. Goldader et al.  2002).  Although the study of the cluster
populations is the primary scientific goal for the ACS imaging, they will be
extremely valuable for understanding the physical conditions that drive luminous
infrared galaxies away from the well-known IRX-$\beta$ correlation - the
correlation between infrared excess and UV spectral slope (see Meurer et al.
1999; Goldader et al. 2002).  ULIRGs fall well off the correlation established
for starburst galaxies, and this is usually interpreted simply as disconnect
between the sources of the observed UV and far-infrared emission caused by dust
obscuration.  However, the physical conditions over which this occurs, and when
this sets in during the merger process is not known.  By combining the ACS,
GALEX and Spitzer imaging data, we will be able to greatly extend the luminosity
range over which the UV (and IR) emission has been mapped, as well as sample
galaxies along the merger sequence.

To understand the true stellar distribution in the dusty circumnuclear regions
in LIRGs, it is important to combine high resolution and long-wavelength
imaging.  The sub-arcsecond HST NICMOS imaging in GOALS will provide us with our
clearest picture of the stellar mass in the LIRG nuclei, and allow us to
accurately de-redden the stellar clusters seen in the ACS data.  With these data
we will be able to search for very small scale, extremely red point sources
which may be buried AGN, and find secondary nuclei hidden from even the deepest
ACS optical observations.  In addition, we will be able to find optically faint
or even invisible clusters around the nuclei, and together with the ACS data,
assemble a true cluster luminosity function unaffected by dust.  The H-band
imaging provides us with the best (existing) near-infrared imaging resolution,
and it samples the peak in the stellar light curve for the old stars which will
dominate the mass in these galaxies.  While the NICMOS F160W and the IRAC
$3.6\mu$m data are both effective tracers of the old stars and stellar mass, the
NICMOS data have a spatial resolution which is an order of magnitude higher - on
the order of 100 pc or better for many of the LIRGs.  By combining the IRAC
imaging of the entire sample with the NICMOS imaging of the luminous GOALS
systems, we will be able to trace the stellar masses from the largest to the
smallest scales.

%something about merger morphologies here

A basic result from the Spitzer imaging in GOALS will be a much better
understanding of the distribution of the dust in mergers.  Since many of the
LIRGs are interacting, yet not fully merged, the IRAC and MIPS images allow us
to separate the contribution from each galaxy to the total far-infrared flux as
measured by IRAS.  They also allow us to separate nuclear from disk emission in
many systems, pinpointing the location of the enhanced infrared luminosity as a
function of merger state.  The MIPS images provide a clean separation between
the warm dust (T$\sim 50-100$K) at $24\mu$m and $70\mu$m, and the cold dust,
which contributes most at $160\mu$m.  One of the most remarkable results from
ISO was the discovery that most of the far-infrared emission in the merging
galaxy NGC 4038/9 (the Antennae) originates from the disk overlap region which
is extremely faint in the UV and optical (Mirabel et al. 1998).  The Spitzer
GOALS images will allow us to find other systems like the Antennae, and
determine the frequency of this completely dust-enshrouded starburst state among
the LIRG population as a whole.

%cosmological nonsense

Taken together, the imaging and spectroscopic data in GOALS will provide the
most comprehensive look at the LIRG population in the local Universe.  By
studying LIRGs from the X-ray through the radio and far-infrared, we can piece
together an understanding of the stars, central black holes, and the
interstellar-medium in some of the most actively evolving galaxies today.  These
observations will be invaluable as a local library with which to interpret the
limited photometric and spectroscopic information garnered from deep surveys of
LIRGs at high-redshift, providing an in-depth glimpse of the co-evolution of
starbursts and black holes at late times, and the role of feedback on the ISM of
merging galaxies.

\section{Sample Definition and Properties}

The IRAS Revised Bright Galaxy Sample (RBGS; Sanders et al. 2003) is a complete
sample of extragalactic objects with IRAS $S_{60} > 5.24$ Jy, covering the full
sky above a Galactic latitude of $|b| > 5$ degrees. The RBGS objects are the
brightest 60-micron sources in the extragalactic sky, and as such they are the
best sources for studying the far-infrared emission processes in galaxies and
for comparing them with observations at other wavelengths. The 629 objects in
the RBGS all have $z < 0.088$, and near-infrared properties spanning a wide
range from normal, isolated gas-rich spirals at low luminosities (L$_{IR} <
10^{10.5}$ \lsun) through an increasing fraction of interacting galaxy pairs and
ongoing mergers among the more luminous LIRGs and ULIRGs. The sample includes
numerous galaxies with optical nuclear spectra classified as starbursts, Type 1
and 2 Seyfert nuclei, and LINERS. The 21 ULIRGs (3\%) and 181 LIRGs (29\%) in
the RBGS form a large, statiscally complete sample of 202 infrared luminous,
local galaxies which are excellent analogs for comparisons with infrared and
sub-mm selected galaxies at high redshift.  These infrared-luminous sources
define a set of galaxies sufficiently large to sample each stage of interaction
and provide a temporal picture of the merger process and it's link to the
generation of far-infrared radiation.  The full sample of LIRGs, along with
their basic properties, is listed in Table 1.  Note that 77 of the LIRG systems
contain multiple galaxies.  From this point on we refer to the 202 LIRGs as
``systems", comprising approximately 291 individual galaxies.  The median distance
to the LIRGs in the GOALS sample is 94.8 Mpc.
Throughout this paper
we adopt $\rm H_0 = 70~km~s^{-1}~Mpc^{-1}$, $\Omega_{vacuum} = 0.72$, and
$\Omega_{matter} = 0.28$.

Note that an update to the cosmological parameters, primarily $\rm H_0 = 70~km~s^{-1}~Mpc^{-1}$ 
(Hinshaw et al. 2009) instead of $\rm H_0 = 75~km~s^{-1}~Mpc^{-1}$, which was used when the IRAS Revised Bright 
Galaxy Sample was compiled (RBGS; Sanders et al. 2003), results in 17 additional RBGS sources 
classified as LIRGs, namely: IRAS F01556+2507
(UGC 01451, log($L_{IR}/L_{\odot}$) = 11.00); F02072-1025 (NGC 0839, log($L_{IR}/L_{\odot}$)=11.01); 
F04296+2923 (log($L_{IR}/L_{\odot}$) = 11.04); F04461-0624 (NGC 1667, log($L_{IR}/L_{\odot}$) = 11.01); 
F06142-2121 (IC 2163, log($L_{IR}/L_{\odot}$) = 11.03); NGC 2341, log($L_{IR}/L_{\odot}$) = 11.17);
F10221-2318 (ESO 500-G034, log($L_{IR}/L_{\odot}$) = 11.01); 
F11122-2327 (NGC 3597, log($L_{IR}/L_{\odot}$) = 11.05); F12112-4659 (ESO 267-G029, log($L_{IR}/L_{\odot}$) = 11.11);
F12351-4015 (NGC 4575, log($L_{IR}/L_{\odot}$) = 11.02);
F14004+3244 (NGC 5433, log($L_{IR}/L_{\odot}$) = 11.02); 
F14430-3728 (ESO 386-G019, log($L_{IR}/L_{\odot}$) = 11.01); F15467-2914 (NGC 6000, log($L_{IR}/L_{\odot}$) = 11.07); 
F17468+1320 (CGCG 083-025; log($L_{IR}/L_{\odot}$) = 11.05); F19000+4040 (NGC 6745, log($L_{IR}/L_{\odot}$) = 11.04); 
F22025+4204 (UGC 11898, log($L_{IR}/L_{\odot}$) = 11.02); F22171+2908 (Arp 278, log($L_{IR}/L_{\odot}$) = 11.02).
Since these galaxies were not part of the original LIRG sample drawn from the RBGS, 
they are not included in GOALS.

\section{GOALS Observations}

\subsection{Spitzer}

\subsubsection{IRAC and MIPS Imaging}

Of the 202 LIRGs in our complete sample, we have obtained images of 175 with
IRAC and MIPS on the Spitzer Space Telescope (PID 3672, PI J. Mazzarella).  The
remaining 27 LIRGs have been observed through other Spitzer GTO and GO programs
and are available in the archive.  The LIRGs were imaged with IRAC at 3.6, 4.5,
5.8, and $8.0\mu$m, in high dynamic range (HDR) mode to avoid saturating the
bright nuclei.  This provided short ($1-2$sec) integrations, in addition to the
longer, primary exposures.  Typically, five exposures of 30sec each were used,
taken in a Gaussian dither pattern.  The LIRGs were imaged with MIPS at 24, 70,
and $160\mu$m using the Photometry and Super Resolution AORs.  Multiple,
$3-4$sec exposures were taken of each source, with two or three mapping cycles
being employed at each wavelength.  Total integration times at each of the three
MIPS wavelengths were 48, 38, and 25sec, respectively.  The IRAC and MIPS photometry
and images are presented in Mazzarella et al. (2009).

\subsubsection{IRS Nuclear Spectroscopy}

As part of GOALS, we have obtained Spitzer IRS (Houck et al. 2004) spectra for 158 LIRGs in Table 1 (PID
30323, PI L. Armus).  Of these, 115 were observed in all four IRS modules
(Short-Low, Long-Low, Short-High, and Long-High), while 43 have been observed in
three or fewer IRS modules in order to complete the existing archival data and
ensure complete coverage for all GOALS targets.

In all cases we have used IRS Staring Mode AORs, employing ``cluster target"
observations for those sources with well separated ( $\Delta r > 10$ arcsec),
nearby interacting companions. Among the 158 LIRGs observed, there were a total
of 202 nuclei targeted. We have targeted secondary nuclei only when the flux
ratio of primary to secondary nucleus (as measured in the MIPS $24\mu$m data) is
less than or equal to five, in order to capture the spectra of the nuclei
actively participating in the far-infrared emission of the system.  Since we
wish to build up a uniform and complete set of nuclear spectra with the IRS, we
have elected to observe all sources in staring mode, even though many LIRGs are
resolved to Spitzer (predominantly at IRAC wavelengths). 

%Accurate coordinates for all the targets are available either from published
%radio maps or the 2MASS Extended Source Catalog (accurate to  0.5 arcsec). The
%targets have nuclear $24\mu$m flux densities (as measured in an 11 arcsec beam
%on the MIPS images that is comparable to the width of the LH and LL slits) that
%range from 0.1-3 Jy. We have divided the sample into three flux density bins,
%$f_{24} < 0.2$ Jy, $0.2 < f_{24} < 0.5$ Jy, and $f_{24} > 0.5$ Jy and developed a set
%of AORs (one for each flux bin, for each set of modules that we are using) to
%consistently deliver high signal-to-noise in the continuum (S/N $ > 20$) in both
%low and high-resolution spectra. For the bright sources, we have used ramp
%durations of 14 sec in SHort-Low (SL) and Long-Low (LL), and 30 and 60 sec in
%Short-High (SH) and Long-High (LH), respectively.  For the faint sources, we
%have used the 60 sec and 30 sec ramps in SL and LL, respectively, and the 120sec
%and 60sec ramps in SH and LH, respectively. In every case we have used at least
%two cycles per nod position. 
%The shortest and longest AORs (for single
%targets) are 11 and 63 mins, respectively. For the proposed SH and LH spectra,
%we are attempting to reach line flux limits of 
%$\sim 1-2 \times 10^{-18}$ W m$^{-2}$
%and equivalent width limits of $0.002-0.003\mu$m across the entire
%sample.  These will allow us to place stringent limits on the high-ionization
%features, the ratios of high to low-ionization lines (e.g., [Ne V]/[Ne II]), and
%the PAH emission feature equivalent widths.

Ramp durations for all modules were selected based upon the measured IRAC and
MIPS nuclear flux densities in order to deliver high signal-to-noise spectra for
all LIRGs.  In all cases we have used at least two cycles per nod position along
the slit, and centered the sources using high accuracy blue IRS peak-up on a
nearby 2MASS star. In order to mitigate against time-varying warm pixels and
measure accurate line equivalent widths in the small high-res slits, we have
obtained background sky observations with matched integration times in the SH
and LH slits. For most sources, backgrounds for the SL and LL slits are obtained
from the non-primary order (sub-slit) when the primary order is on the target.
However, for the (11) targets with IRAC $8\mu$m diameters of  2.0 arcmin or
larger, we have obtained off-source backgrounds, as is done for the high-res
modules.  After background subtraction, residual bad pixels will be removed with
the IRSCLEAN software available from the SSC, and spectra will be extracted
(from 2D to 1D) using the SPICE software.  The first results from the IRS 
spectral survey are presented in Petric et al. (2009).

\subsubsection{IRS Spectral Mapping}

In order to produce spectra of the total mid-infrared emission of the systems,
and to measure variations within the LIRG disks, we are obtaining IRS
low-resolution (SL + LL) spectral maps of 25 LIRGs as part of a Spitzer cycle-5
GO program (PID 50702, PI L. Armus).  The target LIRGs were chosen to span a
large range in morphological properties (isolated through late stage
interactions) while being well-resolved in our IRAC $8\mu$m images (to ensure
adequate area for constructing spatially-resolved spectral maps).  Distances to
the mapped LIRGs range from 45.5 to 105.8 Mpc, implying projected spatial
resolutions at $6\mu$m of 0.7 to 1.8 kpc.  (3.6 arcsec, or two SL pixels).  Ten
of these systems have two well separated galaxies requiring individual maps, and
therefore there are 35 individual maps in the program.  In addition,
there are 17 LIRGs with IRS spectral mapping data
in the Spitzer archive.

All LIRGs will be observed with the SL and LL slits.  There are typically 
$25-45$ steps in each SL map, and $7-13$ steps in each LL map, to cover areas in
both slits that are $0.5-2$ sq. arcminutes on each source.  In all cases, step
sizes perpendicular to the slits are 1/2 the width of the slit.  The maps are
designed to cover the main bodies of the galaxies, but not the tidal tails (for
those LIRGs with tails).  The SL exposures are all 60 sec per position, while
the LL exposures are 30 sec per position.  IRS peakups are not used for the
mapping observations.  The AORs, which include the SL and LL maps, range in time
from $0.8 - 6.0$ hrs each, including overheads.  The spectral cubes will be
assembled, and 1D spectra extracted using the CUBISM software package available
from the SSC web site (Smith et al. 2007).

\subsection{Hubble}

\subsubsection{ACS WFC Imaging}

The {\it HST} observations are presently composed of three campaigns. The first
component consists of Advanced Camera for Surveys (ACS) observations of all 
88 LIRGs with $L_{\rm IR} > 10^{11.4}$ L$_\odot$.
Many of these LIRGs have tidal structure or widely separately nuclei on scales
of tens to a hundred arc-seconds.  Thus the Wide Field Channel (WFC), with its
$202\arcsec \times 202\arcsec$ field of view, was selected to capture the
full extent of each interaction in one {\it HST} pointing. Each LIRG was imaged
with both the F435W and F814W filters, with integration times of 20 and 10
minutes, respectively. A total of 88 orbits of ACS/WFC data were obtained (PID
10592, PI A. Evans).  The ACS imaging atlas and initial photometric measurements
are presented in Evans et al. (2009).

\subsubsection{NICMOS Imaging}

The second {\it HST} campaign was  designed to provide high-resolution
near-infrared observations to recover nuclear structure obscured from view at
optical wavelengths. Observations with the Near-Infrared Camera and Multi-Object
Spectrometer (NICMOS) of 59 LIRGs in the ACS-observed sample are underway.
The data are being collected using camera two (NIC2) with a field of view of
$19.3\arcsec \times 19.5\arcsec$, using the F160W filter. For most of the
observations, one LIRG is observed per orbit, however, there are many cases in
which the galaxy pairs are too widely separated to be observed by {\it HST}
in a single orbit. These new data are complemented by pre-existing archival
NICMOS data of an additional 29 GOALS LIRGs, thus providing a complete set of
NIC2/F160W observations of the 88 LIRGS in the ACS-observed sample. A total of
76 orbits of NICMOS data are being obtained (PID 11235, PI J. Surace).

\subsubsection{ACS SBC Imaging}

The third {\it HST} campaign is designed to obtain near- and far-UV data of a
subset of 22 of the LIRGs previously observed with the ACS/WFC that possess the
highest number of luminous star clusters within their inner 30$\arcsec$ (i.e.,
the field-of-view of the Solar Blind Channel, SBC, of ACS). Five of the 22 LIRG
systems are in widely separated pairs that require two pointings, so there
are 27 galaxies being observed in total.  There are
three additional LIRGs with adequate far-UV imaging in the HST archives, 
bringing the total
number of LIRG systems in GOALS with far-UV HST data to 25.  The UV observations
are being obtained by using the ACS/SBC and WCPC2 Planetary Camera (PC) in
combination with the F140LP (far-UV) and F218W (near-UV) filters, respectively.
The program (PID 11196, PI A. Evans) consists of three orbits per galaxy, with
one orbit devoted to ACS/SBC (four individual integrations in a box pattern),
and two orbits devoted to WFPC2 (four indivdual integrations total over both
orbits taken in a box pattern).  A total of 81 orbits of UV imaging data are
being obtained.

\subsection{Chandra}

Currently, the Chandra X-ray Observatory observations of the GOALS
targets focuses on those LIRGs with infrared luminosity larger than
$10^{11.73}$L$_{\odot}$. We observed 26 targets during the Chandra Cycle-8 (PID
8700551, PI D. Sanders).  The X-ray imaging observations were carried out using
the ACIS-S detector in VFAINT mode with a 15 ks exposure for each target.
Combined with the previously observed 18 objects from the archive, the current
sample consists of 44 objects. The exposure time for the objects from the
archive ranges from 10 ks to 160 ks. ESO203-IG001 is the only galaxy which was
not detected. The products available for each source are the contours of a
Chandra full-band image overlaid onto the HST/ACS I-band image; X-ray images in
the full (0.4-7 keV), soft (0.5-2 keV), and hard (2-7 keV) band images;
azimuthally-averaged radial surface brightness profiles in the soft and hard
bands; an energy spectrum when sufficient counts are available (typically more
than 100 counts). Other basic information is presented in 
Iwasawa et al. (2009a,b), and Teng et al. (2009), including detected source counts, soft and hard
band fluxes and luminosities, the X-ray hardness ratio (as a guide of a rough
estimator of the spectral shape), and a measure of spatial extension of the soft
X-ray emission.

\subsection{GALEX}

While the LIRGs are, by definition, infrared luminous, they often do have
measurable UV fluxes.  In fact, due to variations in the spatial distribution
of the obscuring dust, a number of well-known (U)LIRGs in GOALS (e.g. Mrk 171, 
Mrk 266, Mrk 231, Mrk 273, Mrk 617, and Mrk 848) were first identified as 
starburst galaxies and AGN by the Markarian survey of UV-excess galaxies -- a
spectroscopic objective-prism survey conducted years before the IRAS survey
(e.g., see Mazzarella \& Balzano 1986).
Of the 202 LIRGs in GOALS, 145 have been observed with
the GALEX telescope in the near-UV (2271 \AA) and far-UV (1528 \AA) filters.  Of
these, 124 have high signal-to-noise detections in both filters.  Most of
these images were taken as part of the All sky Imaging Survey (AIS) or the
Nearby Galaxy Survey (NGS - see Gil de Paz et al. 2007), with 51 being observed
as part of GOALS (GI1-013, PI J. Mazzarella and GI5-038, PI J. Howell).  The AIS integration times are
short (100 sec), while the NGS and GOALS observations are long, typically a few 
ksec in duration.  Details of the GALEX observations and the UV photometry
are presented in Howell et al. (2009).

\section{GOALS Data Products}

On regular, approximately six month, intervals, Spitzer images and spectra will
be delivered to the Spitzer Science Center and made public through their Legacy
program web pages.  The first delivery is now available through the Spitzer
Legacy web pages at the Spitzer Science Center
(http://ssc.spitzer.caltech.edu/legacy/all.html), and the Infrared Science
Archive (IRSA) at the Infrared Processing and Analysis Center
(http://irsa.ipac.caltech.edu).  For GOALS, the delivered products include, (1)
IRAC image mosaics in all four IRAC bands.  Images are single-extension FITS
files with a pixel scale of 0.6 arcsec/pixel.  (2) MIPS image mosaics in all
three MIPS bands.  Images are single-extension FITS files, with
wavelength-dependent pixels cales -- 1.8 arcsec/pixel at $24\mu$m, 4.0
arcsec/pixel at $70\mu$m, and 8.0 arcsec/pixel at $160\mu$m.  (3) IRS nuclear
spectra in all four low and high-resolution modules.  Spectra are delivered in
ASCII (IPAC table) format, similar to those produced by the IRS pipelines.  (4)
Spatial profiles at two positions along the IRS Short-Low slit (the two nod
positions), in the $8.6\mu$m PAH, $10\mu$m continuum, and $12.8\mu$m [Ne II]
fine-structure emission line.  Profiles are provided in ASCII format, in units
of e$^{-}$/sec for each pixel.  Also included in each file are the profiles at
the same wavelengths for an unresolved star.  The IRAC images are corrected for
``banding", ``muxbleed", and ``column pulldown" effects, as described in the
IRAC Data Handbook.  The MIPS images are corrected for latents, weak jailbars,
and saturation effects, where possible.  The IRAC and MIPS mosaics were
constructed using MOPEX to align, resample and combine the data.  The IRS
spectra are extracted from the two-dimensional BCD pipeline products using the
SPICE package.  A standard ``point source" extraction has been used for all
sources.  The two nod positions and multiple exposures (``cycles") are used to
remove residual cosmic rays and/or bad pixels in the spectra.

\section{The Luminous Infrared Galaxy VV 340}

As an example of the utility of the multi-wavelength GOALS data, we
present images and spectra of the  LIRG, VV 340 (IRAS F14547+2449;
Arp 302).  VV 340 consists of two large, spiral galaxies, one
face-on and one edge-on, separated by approximately 40 arcsec (27.3 kpc).
The two spirals are apparently in the early stages of a merger 
(Bushouse \& Stanford 1992, Lo, Gao \& Gruendl 1997).  This system
has an infrared luminosity of $5\times10^{11}$L$_{\odot}$ (Sanders
et al. 2003).

\subsection{Observations and Data Reduction}

IRAC observations of VV 340 were performed on 17 July 2005 (PID 3672).
The data were obtained in high dynamic range mode, with a 30 sec frame
time and a five position Gaussian dither pattern.
MIPS observations of VV 340 were performed on 25 January 2005 (PID 3672).
The data were obtained in photometry mode, using the small field size
and an integration time of 3 sec in all three (24, 70, $160\mu$m) filters.
One cycle was used for 24 and $70\mu$m, with four cycles being used at $160\mu$m.

IRS observations of the two VV 340 nuclei were performed on 17 March 2007 (PID
30323).  The data were obtained in staring mode, using the cluster option.  A
moderate accuracy blue peakup was done on a nearby 2MASS star in order to center
the IRS slits on the northern and southern nuclei.  On the sky, the IRS
Long-low, Short-low, Short-high, and Long-high slits had projected position
angles of 94.1, 177.8, 134.4, and $49.5^{o}$, respectively.  The Long-low,
Short-low, Short-high and Long-high observations had ramp times of 60, 30, 120
and 240 seconds, respectively.  The high-res observations also had corresponding
offset sky observations with the same ramp times.  The data were reduced using
the S15 version of the IRS pipeline at the SSC, and the spectra were extracted
using the SPICE software package using the standard, point source extraction
aperture.

GALEX near-UV and far-UV observations were performed on 19 May 2005 and 29 April
2007 (PID 13), for a total of 6028 sec and 3042 sec, respectively.  VV 340 was
observed with the Chandra X-ray Observatory on 17 December 2006 (PID 8700551).
An exposure of 14.9 ks was used, yielding 285 net counts.

HST ACS observations of VV 340 were performed on 7 January 2006 (PID 10592).
The data were obtained using the WFC in ACCUM mode, in the F435W and F814W filters.
Three, 420 sec exposures were taken in the F435W filter and two, 360 sec exposures
were taken in the F814W filter using the LINE dither routine.  The WFC field of 
view is $202\times 202$ arcsec, and the pixel size is 0.05 arcsec.  The data were
run through the standard pipeline processing, which removes instrumental signatures
subtracts a dark frame, flat fields the images, removes geometric distortion and 
applies a flux calibration.  Additional cosmic rays were identified and removed
using the lacos\_im and jcrrej2.cl routines in IRAF (Rhoads 2000; Van Dokkum 2001).

\subsection{Results \& Discussion}

The Chandra, GALEX, HST, and Spitzer images of VV 340 are shown in Fig. 1.
A false-color image made from the ACS F435W and F814W data is shown in Fig.
2.  The thick, flaring dust lane makes VV 340 N much fainter in the UV than
VV 340 S,  but the disk and bulge are prominent in the IRAC images.  Using
the ACS images, we have detected 173 unresolved ``clusters" in the VV 340
system, with most of these (159, or 92\%) being in the spiral arms of VV 340
South.  The apparent magnitudes range from $21-27$ mag in B, with absolute
magnitudes ranging from 9 to -13 mag, and $B-I$ colors of $0.5-2$ mag.
These colors are typical for the GOALS LIRGs as a class (Vavilkin et al.
2009).  Assuming a typical cluster mass of $10^6$M$_{\odot}$, the ages of
these clusters are $\le 10^{8}$ yrs, uncorrected for reddening.  As a
whole, the observed clusters in VV 340 South account for 3\% of the total
B-band light.

The total far-infrared flux densities for the VV 340 system, as measured
with MIPS, are 0.43, 9.38, and 15.73 Jy at 24, 70, and $160\mu$m,
respectively.  The northern galaxy dominates at 24, 70 and $160\mu$m,
accounting for approximately 80, 82, and $\sim95$\%, respectively, of the total
emission from the system (the $160\mu$m ratio is uncertain because of the
large size of the PSF).  For reference, the IRAS 25, 60, and $100\mu$m flux
densities for the entire VV 340 system are 0.41, 6.95, and 15.16 Jy (Sanders
et al. 2003).  It is interesting to note that the ratio of far-infrared
fluxes implies that VV 340 South, by itself, is not a LIRG (log L$_{IR} \sim
10.71$L$_{\odot}$).

The IRAC $3.6-4.5$ and $5.8-8$ colors, as measured in 10 kpc radius
apertures centered on the two nuclei are 0.19 and 1.97 mags for VV 340
North, and 0.04 and 1.96 mags for VV 340 South, respectively.  The nuclear
(2 kpc radius aperture) colors are slightly redder, being 0.29 and 2.06
mags for VV 340 North, and 0.06 and 2.16 mags for VV 340 South,
respectively.  Both the large and small aperture measurements place both
galaxies in VV 340 outside of the AGN ``wedge" in the IRAC color-color
diagram of Stern et al. (2005).

The Spitzer IRS low resolution spectra of the northern and southern nuclei are
shown in Fig. 3.  VV 340 N has a steeper spectrum and a deeper silicate
absorption ($\tau_{9.7} \sim 1.3$ for VV 340 N).  The PAH emission features are
prominent in both galaxies, although the $17\mu$m feature appears much stronger,
with respect to the continuum and the fine structure lines, in VV 340 S.  Both
[Ne V] 14.3 and $24.3\mu$m lines are seen in the high resolution spectrum of
VV 340 N, as is the [O IV] $25.9\mu$m line.  The line fluxes are $1.2\times
10^{-17}$ W m$^{-2}$, $2.3\times 10^{-17}$ W m$^{-2}$, and $12.3\times 10^{-17}$
W m$^{-2}$, repectively for the [Ne V] 14.3, $24.3\mu$m, and [O IV] $25.9\mu$m
lines in VV 340 N.  A portion of the Short-high spectra of VV 340 N and S
surrounding the location of the [Ne V] $14.3\mu$m emission line is shown in Fig.
4.  The [Ne II] $12.8\mu$m line has a flux of $23.4\times 10^{-17}$ W m$^{-2}$ in
VV 340 N, and $6.1\times 10^{-17}$ W m$^{-2}$ in VV 340 S.  In VV 340 N, where
the $6.2\mu$m PAH EQW is $0.48\mu$m, the [Ne V]/[Ne II] and [O IV]/[Ne II] line flux
ratios are consistent with the presence of a weak AGN, contributing less than
$10-15$\% of the IR emission, based on scaling from IRS spectra of local AGN 
and starburst nuclei (see Armus et al. 2007 and references therein).  
Since the diagnostic ratios are calculated from
the nuclear spectrum, this is an upper limit to the contribution from an AGN to
the global IR emission in this source.  Alternatively, the coronal-line region could be
sitting behind at least 50 mag of visual extinction, assuming an intrinsic
[Ne V]/[Ne II] flux ratio of unity.  This is much larger than that implied by the
optical depth at $9.7\mu$m (A$_{V} \sim 22$ mags) for VV 340 N.  There is no
[Ne V] detected from VV 340 S ([Ne V]/[Ne II] $ < 0.04$), and the $6.2\mu$m PAH EQW
is $0.53\mu$m, comparable to pure starburst galaxies (Brandl et al. 2006).  The
lack of [Ne V] emission and the strong PAH EQW both suggest that the southern
nucleus has no detectable AGN in the mid-infrared.

The H$_{2}$ S(2), S(1), and S(0) rotational lines at 12.28, 17.03, and
28.22$\mu$m are detected in VV 340 N with line fluxes of $5.67\times10^{-17}$ W
m$^{-2}$, $4.56\times10^{-17}$ W m$^{-2}$, and $1.3\times10^{-18}$ W m$^{-2}$,
respectively.  In VV 340 S, only the S(2) and S(1) lines are seen, with fluxes
of $4.4\times10^{-18}$ W m$^{-2}$, and $1.0\times10^{-17}$ W m$^{-2}$,
respectively.  Fits to the emission lines imply masses and temperatures for the
warm gas components of $1.73\times 10^{7}$ M$_{\odot}$ at 530K in VV 340 N, and
$7.8\times 10^{6}$ M$_{\odot}$ at 310K in VV 340 S.  The nuclear warm gas
fractions (warm/cold) are then $\sim 3\times 10^{-4}$ and $\sim 7\times 10^{-4}$
in VV 340 N and VV 340 S, respectively, using the masses of cold H$_{2}$ derived
by Lo, Gao \& Gruendl (1997).  However, the large extent of the CO emission, 23
kpc in VV 340 N and 10 kpc in VV 340 S (much larger than the IRS slit widths)
suggests that these ratios are probably lower limits for the nuclei.  The large
extent of the CO, the relatively low infrared luminosity to H$_{2}$ mass ratio
in both galaxies, and the regular kinematics led Lo, Gao \& Gruendl to suggest
that the VV 340 system is in an early interaction, pre-starburst, phase.

VV 340 has a total (North plus South) UV flux of 17.55 mag and 16.64 mag in
the far and near-UV GALEX filters, respectively.  The pair has an apparent
infrared to UV flux ratio (the infrared excess, or IRX), of 81.3.  The
measured UV slope, $\beta$, is $-0.38$ on the GALEX system (see Kong et al. 2004), 
placing VV 340 well above the
fit to local starburst galaxies (Meurer et al. 1999; Goldader et al. 2002)
by nearly an order of magnitude.  As is obvious from Fig.1, most of the UV
emission comes from VV 340 South, while most of the IR emission comes from
VV 340 North.  If we were to place the galaxies on an IRX-$\beta$ plot
individually, the Northern galaxy would have an IRX $\sim 398$ and a $\beta
\sim -0.39$, while the Southern galaxy would have an IRX $\sim 17$ and a
$\beta \sim -0.44$.  This would place the Southern source close to the
starburst correlation, but the Northern source much farther off the
correlation than the system as a whole.  A full discussion of where the
LIRGs fall on the the IRX-$\beta$ plot is given in Howell et al. (2009).

VV 340 is detected at both 1.49 and 4.85 GHz with the VLA (Condon et al. 1990;
1995).  While both VV 340 N and S are resolved in the VLA C-array data, the
A-array data also reveals a nuclear component and two ``hotspots" at radii of
about five arcsec, oriented N-S, along the edge-on disk of VV 340 N.  VV 340 N
and VV 340 S have 1.49 GHz flux densities of 86, and 12 mJy, respectively, and
4.85 GHz flux densities of 30 and 3.5 mJy, respectively.  The system has a
logarithmic far-infrared to radio flux ratio, $q$ (Helou et al. 1985) of
$q=2.02$~dex.  This is within $1.5\sigma$ of the standard value for star-forming
galaxies ($q=2.34 \pm 0.26$~dex; Yun, Reddy \& Condon 2001).  VV 340 N and VV 340
S have $q$ values of 2.02 and 2.00~dex, respectively. VV 340 S has a slightly
steeper radio spectral index ($\alpha = -1.0$) than does VV 340 N ($\alpha =
-0.8$), the latter being comparable to the standard value seen in star-forming
galaxies (Condon 1992).

Chandra X-ray imaging of VV 340 is shown in Figs. 5 \& 6.  While both
galaxies are detected, the emission from the southern galaxy is much
weaker, and more diffuse.  The northern galaxy is clearly resolved in the
soft bands ($0.5-2$ keV), with a size of nearly 25 arcsec (17 kpc), but shows a hard
($2-7$ keV) X-ray point source coincident with the nucleus.  The hard X-ray
image also shows two faint blobs to the north and south, aligned with the
radio structure (Condon et al. 1990).

The X-ray spectrum of VV 340 N (Fig. 7) is dominated by emission-line rich, extended
soft X-ray emission but an excess at high energies above 4 keV is
evident. In Fig. 7 we compare the Chandra spectrum of VV 340 N to that
of the starburst-dominated LIRG VII Zw 31 (Iwasawa et al. 2009b).  VII Zw 31
shows no hard-band excess above 4 keV.
The hard-band excess in VV 340 N peaks at approximately 6.4 keV in the rest frame,
suggesting a pronounced Fe K line. With the limited detected counts, the
exact spectral shape cannot be constrained.  However, the most likely
interpretation for this hard excess is the emission from a heavily absorbed
active nucleus with an absorbing column density ($N_{\rm H}$) of $10^{24}$
cm$^{-2}$ or larger. Note, this is significantly larger than that implied
by the silicate optical depth at $9.7\mu$m.
If the observed emission is reflected light from a Compton thick
AGN, the intrinsic luminosity could be as high as
$2\times 10^{41}$ erg s$^{-1}$ in the 2-7
keV band.  There is also evidence for the 1.8 keV Si line, which could
have a contribution from AGN photoionization, although it could also signal
the presence of a large number of core-collapse SNe.

Taken together, the Spitzer, Chandra, GALEX, HST, and VLA data suggest that the
interacting system VV 340 is composed of two very different galaxies.  The
edge-on disk of VV 340 North hides a buried AGN seen in the mid-infrared and
X-rays, although the apparent contribution of this AGN to the total energy of
the galaxy is low.  In contrast, VV 340 South is a face-on starburst galaxy that
generates an order of magnitude less far-infrared flux, but dominates the
short-wavelength UV emission from the system.  VV 340 is a LIRG because of the
enhanced emission coming from VV 340 N alone.  VV 340 seems to be an excellent
example of a pair of interacting galaxies evolving along different paths, or at
different rates, implying that the details of the interaction can produce LIRGs
whose global properties mask the true nature of the emission.  This is
particularly relevant for merging galaxies viewed at high redshift.

\acknowledgements

The Spitzer Space Telescope is operated by the Jet Propulsion Laboratory,
California Institute of Technology, under NASA contract 1407.  This research has
made use of the NASA/IPAC Extragalactic Database (NED) which is operated by the
Jet Propulsion Laboratory, California Institute of Technology, under contract
with the National Aeronautics and Space Administration.  Based on observations
made with the NASA Galaxy Evolution Explorer.  GALEX is operated for NASA by the
California Institute of Technology under NASA contract NAS5-98034.  TV, ASE and
DCK were supported by NSF grant AST 02-06262 and by NASA through grants
HST-GO10592.01-A and HST-GO11196.01-A from the Space Telescope Science Institute, which is operated by the Association of Universities for Research in
Astronomy, Inc., under NASA contract NAS5-26555. TV and HI acknowledge support from
the Spitzer Graduate Student Fellowship Program.  The authors wish to thank an anonymous referee for
suggestions which improved the manuscript.

\begin{figure*}[multilampic]
%\plotone{VV3401.ps}
%\caption{Multi-wavelength images of VV 340.  From upper left to lower right, these are (a) GALEX FUV, 
%(b) GALEX NUV, (c) HST ACS F435W, (d) HST ACS F814W), (e) Spitzer IRAC 3.6$\mu$m, (f) Spitzer IRAC
%4.5$\mu$m, (g) Spitzer IRAC 5.8$\mu$m, and (h) Spitzer IRAC 8$\mu$m.}
%\plotone{figv54.ps}
%\caption{Multi-wavelength images of VV 340.  From upper left to lower right, these are 
%(a) HST ACS F435W, (b) HST ACS F814W), (c) Spitzer IRAC 3.6$\mu$m, (d) Spitzer IRAC
%4.5$\mu$m, (e) Spitzer IRAC 5.8$\mu$m, (f) Spitzer IRAC 8$\mu$m., (g) Spitzer MIPS $24\mu$m, and 
%(h) VLA 1.49 GHz.  In all cases a projected linear scale of 5 kpc at the distance of VV 340 is
%indicated by a bar in the lower left.}
\begin{center}
\resizebox{7.5in}{!}{
\plotone{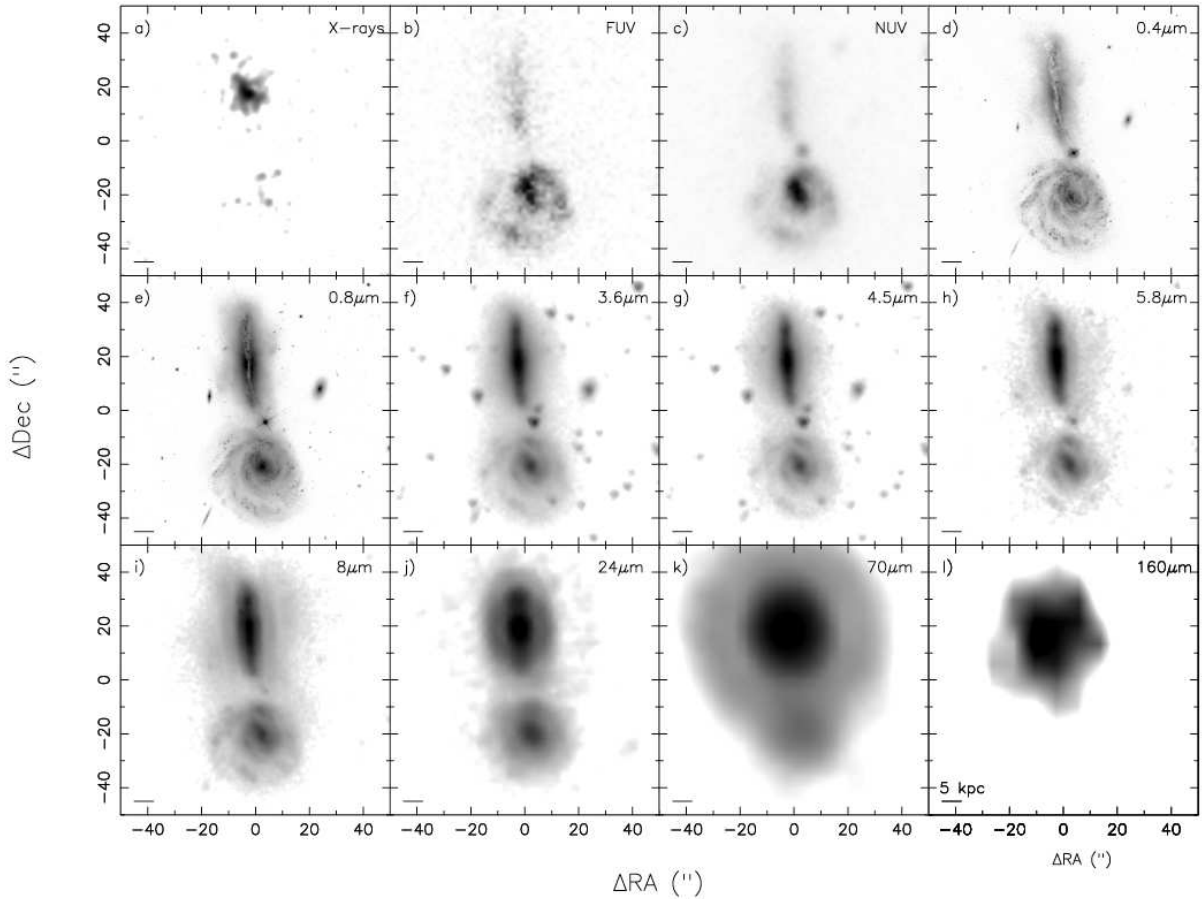}}
\end{center}
\caption{Multi-wavelength images of VV 340.  From upper left to lower right, these are
(a) Chandra $0.4-8$ keV, (b) GALEX FUV, (c) GALEX NUV, (d) HST ACS F435W, 
(e) HST ACS F814W), (f) Spitzer IRAC 3.6$\mu$m, (g) Spitzer IRAC
4.5$\mu$m, (h) Spitzer IRAC 5.8$\mu$m, (i) Spitzer IRAC 8$\mu$m., (j) Spitzer MIPS $24\mu$m, 
(k) Spitzer MIPS $70\mu$m, and (l) Spitzer MIPS $160\mu$m.  In all cases a projected linear scale of 5 kpc at the distance of VV 340 is
indicated by a bar in the lower left.}
\end{figure*}

\begin{figure*}[hstcolorpic]
\begin{center}
\resizebox{5.5in}{!}{
\plotone{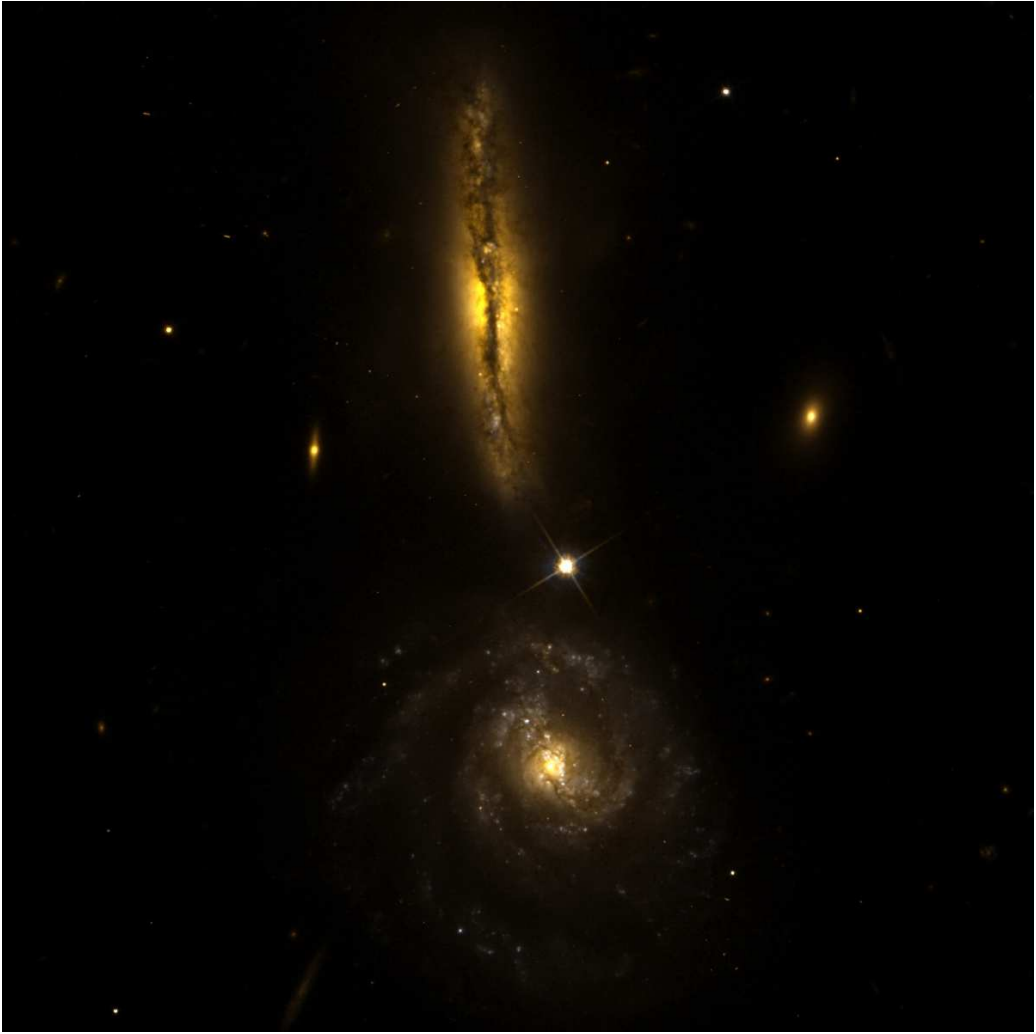}
%\plotone{VV340mini2-log-cmyk.eps}
}
\end{center}
\caption{False color image of VV 340 (UGC 9618) made from the ACS F435W (blue) and F814W (red) data.  The
image is approximately 100 arcsec (68 kpc) on a side.  North is up and East is to the left.}
\end{figure*}

\begin{figure}[IRSlowres1]
\begin{center}
%\resizebox{5.5in}{!}{
\plotone{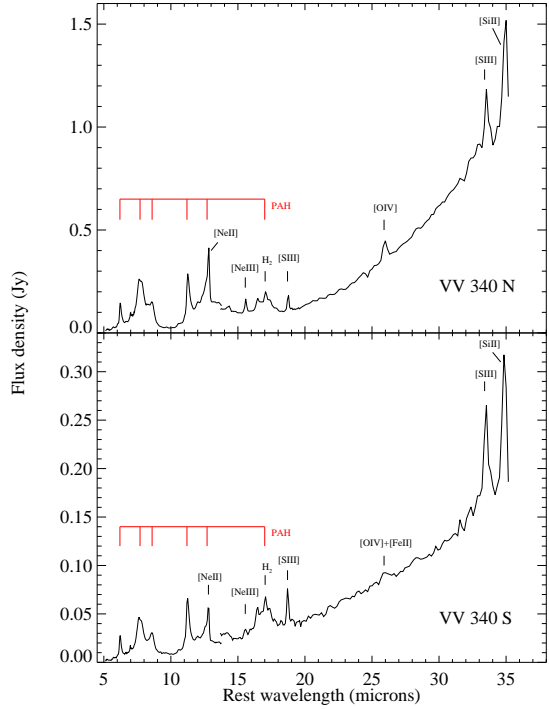}
%}
\end{center}
%\plotfiddle{VV340_spectra2.eps}{5in}{0.0}{500.}{500.}{-50}{500}
\caption{IRS low resolution spectra of the nucleus of VV 340 North (above), and South (below).  
Prominent emission features are marked.  The offset at $14\mu$m in the spectrum of VV 340 S
is real, and represents a flux difference in the SL and LL spectra for this source, most 
likely due to significant extended emission in the LL slit.}
\end{figure}

\begin{figure}[IRShighres1]
\begin{center}
%\resizebox{4.5in}{!}{
\plotone{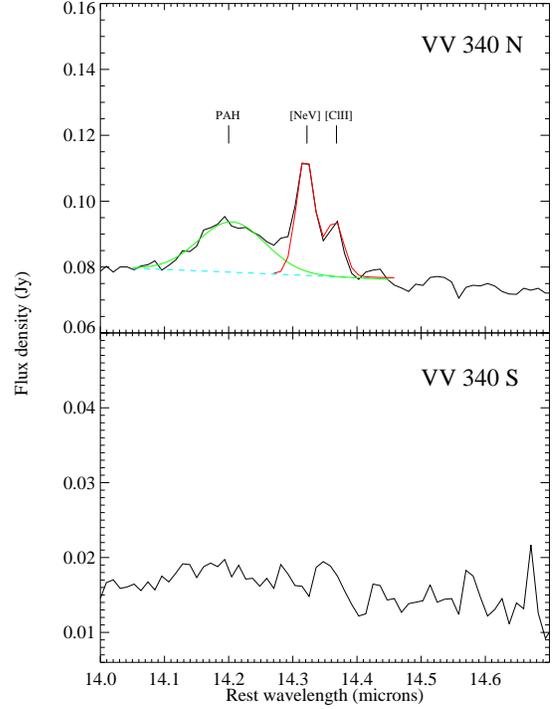}
%}
\end{center}
\caption{IRS Short-high spectra of the nucleus of VV 340 North (above), and South (below) from 
$14-14.7\mu$m in the rest frame.  Fits to the PAH $14.22\mu$m emission feature, as well as the 
[Ne V] $14.322\mu$m, and [Cl II] $14.368\mu$m emission lines indicated in green and red, respectively 
in VV 340 N, 
as is the underlying local continuum (blue dashed line).  The lines are not detected in VV 340 S.}
\end{figure}

\begin{figure}[xrayonHST]
\begin{center}
%\resizebox{4.5in}{!}{
\plotone{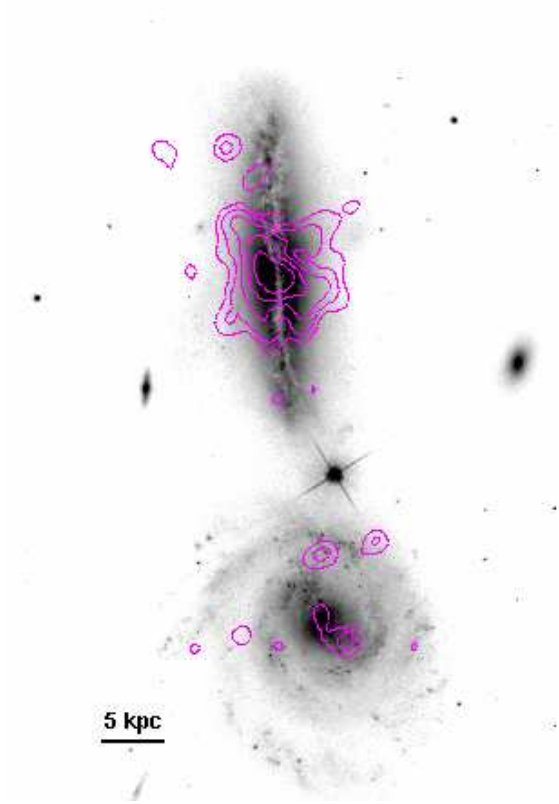}
%}
\end{center}
\caption{Soft ($0.2 - 5$ keV) X-ray image of VV 340 taken with the Chandra X-ray Observatory (contours),
overlayed on the HST ACS F435W image (greyscale).  North is up, and East is to the left.}
\end{figure}

\begin{figure}[xraysofthard]
\begin{center}
%\resizebox{4.5in}{!}{
\plotone{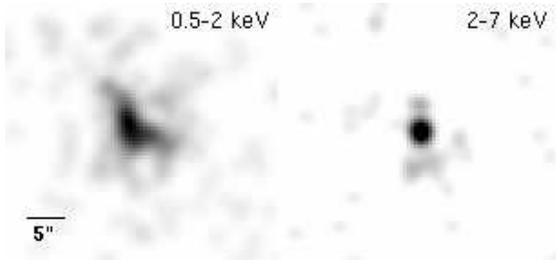}
%}
\end{center}
\caption{The soft (left) and hard (right) X-ray images of VV 340 North taken 
with the Chandra X-ray Observatory.  The images have been smoothed with a
Gaussian kernel with $\sigma =2$ pixels.  The orientation of the image is
North up and East to the left.}
\end{figure}

\begin{figure}[xrayspectrum]
%\plotone{vv340cxospec_rot.ps}
\begin{center}
%\resizebox{4.5in}{!}{
\plotone{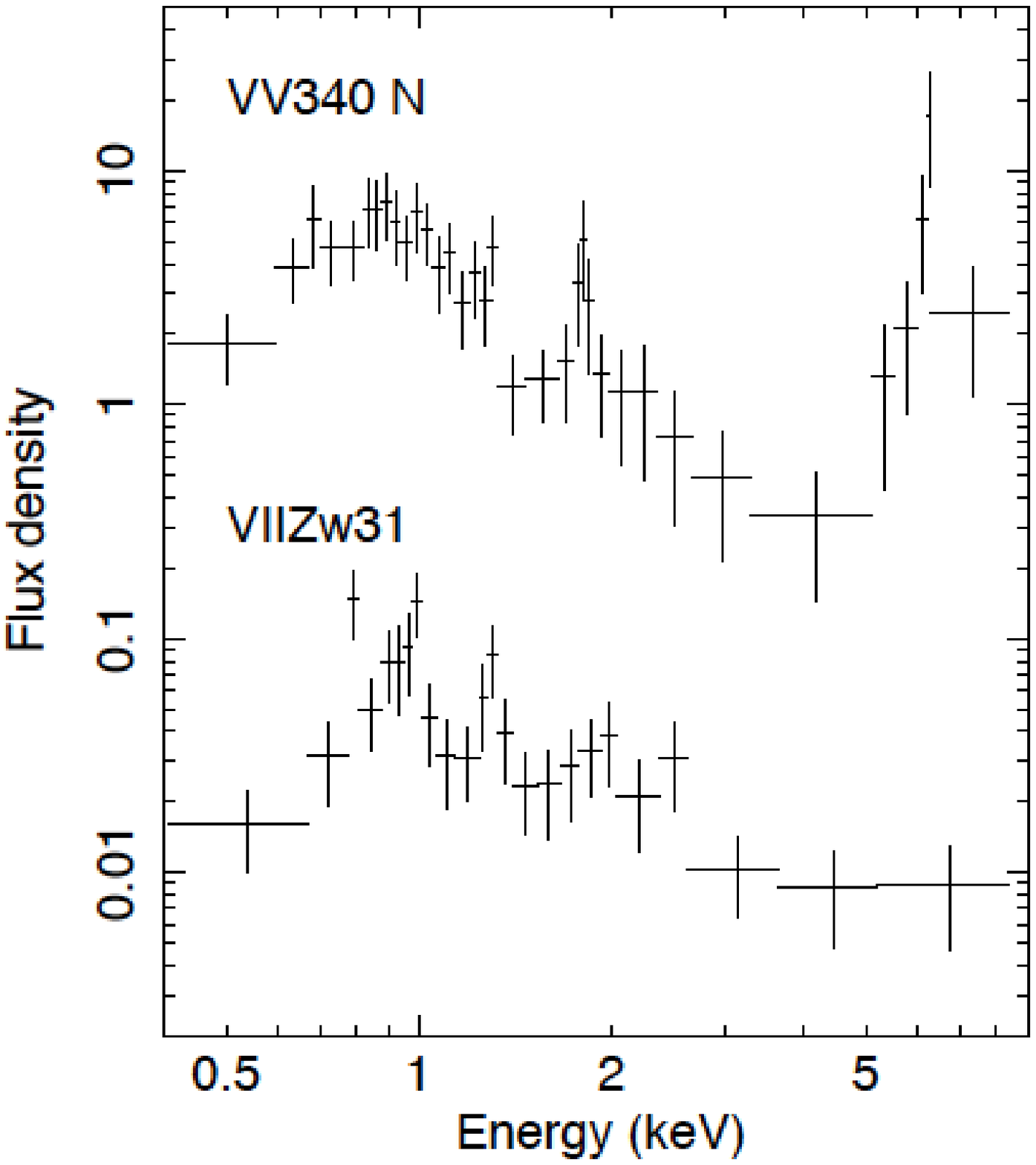}
%}
\end{center}
\caption{The X-ray spectrum of VV 340 North obtained from the ACIS-S.  For display
purposes, the original data have been rebinned so that each spectral bin contains
at least five counts.  The vertical axis is in units of $10^{-14}$erg cm$^{-2}$ s$^{-1}$ keV$^{-1}$.
Also included for comparison is the ACIS-S spectrum of VII Zw 31
from Iwasawa, et al. (2009b), scaled by 0.02, which is an example of a starburst-dominated LIRG.  The
hard X-ray excess in VV 340 North is evident at energies above about 4 keV.}
\end{figure}

%\clearpage
\references

%\reference{} Armus, L., Heckman, T.M., Weaver, K.A., \& Lehnert, M.D. 1995, ApJ, 445, 666.

\reference{} Armus, L., Heckman, T.M., \& Miley, G.K. 1987, AJ, 94, 831.

\reference{} Armus, L., Heckman, T.M., \& Miley, G.K. 1989, ApJ, 347, 727.

\reference{} Armus, L., et al. 2004, ApJ Suppl., 154, 178.

\reference{} Armus, L., et al. 2006, ApJ, 640, 204.

\reference{} Armus, L., et al. 2007, ApJ, 656, 148.

\reference{} Barnes, J.E., \& Hernquist, L. 1992, ARA\&A, 30, 705.

\reference{} Blain, A.W., Smail, I., Ivison, R.J., Kneib, J.-P., \& Frayer, D.T. 2002, PhR, 369, 111.

\reference{} Brandl, B.R., et al. 2006, ApJ, 653, 1129.

\reference{} Bushouse, H.A., \& Stanford, S.A. 1992, ApJS, 79, 213.

\reference{} Chapman, S.C., Blain, A.W., Smail, I., \& Ivison, R.J. 2005, ApJ, 622, 772.

\reference{} Caputi, K.I., et al. 2007, ApJ, 660, 97.

\reference{} Condon, J.J., Helou, G., Sanders, D.B. \& Soifer 1990, ApJS, 73, 359.

\reference{} Condon, J.J. 1992, Annual Reviews of Astronomy \& Astrophysics, 30, 575.

\reference{} Condon, J.J., Anderson, E., \& Broderick, J.J. 1995, AJ, 109, 2318.

\reference{} Dahlem, M., Weaver, K.A., \& Heckman, T.M. 1998, ApJS, 118, 401.

\reference{} Desai, V., et al. 2007, ApJ, 669, 810.

\reference{} Draine, B.T. \& Li, A. 2001, ApJ, 551, 807.

\reference{} Dickinson, M., et al. 2003, in ``The Mass of Galaxies at Low and High Redshift" ESO Astrophysics Symposia, edited by R. Bender and A. Renzini, Springer-Verlag, p. 324.

\reference{} Dunne, L., et al. 2000, MNRAS, 315, 115.

\reference{} Elbaz, D., et al. 2002, A\&A, 384, 848.

%\reference{} Evans, A.S., Mazzarella, J.M., Surace, J.A., \& Sanders, D.B. 2002, ApJ, 580.

\reference{} Evans, A.S., Mazzarella, J.M., Surace, J.A., Frayer, D.T., Iwasawa, K., \& Sanders, D.B. 2005, ApJS, 158, 197.

\reference{} Evans, A.S., et al. 2008, ApJ, 675, 69L.

\reference{} Evans, A.S., et al. 2009, in preparation.

\reference{} Fabbiano, G., Heckman, T.M., \& Keel, W.C. 1990, ApJ, 355, 442.

\reference{} Ferrarese, L., \& Merritt, D. 2000, ApJ, 539, 9L.

\reference{} Gao, Y., \& Solomon, P.M. 2004, ApJ, 606, 271.

\reference{} Gebhardt, K., et al. 2000, ApJ, 539, 13L.

\reference{} Genzel, R., Lutz, D., Sturm, E., Egami, E., Kunze, D., 
et al. 1998, ApJ, 498, 589.

\reference{} Genzel, R., Tacconi, L.J., Rigopoulou, D., Lutz, D., \& Tecza, M. 2001, ApJ, 563, 527.

\reference{} Gil de Paz, A., et al. 2007, ApJS, 173, 185.

\reference{} Goldader, J.D., et al. 2002, ApJ, 568, 651.

\reference{} Grimm, H.-J., Gilfanov, M., \& Sunyaev, R. 2003, MNRAS, 339, 793.

%\reference{} Heckman, T.M., Armus, L., \& Miley, G.K. 1987, AJ, 93, 276.

\reference{} Heckman, T.M., Armus, L., \& Miley, G.K. 1990, ApJ Suppl., 74, 833.

\reference{} Helou, G., Soifer, B.T., \& Rowan-Robinson, M. 1985, ApJ, 298, 7L.

\reference{} Hinshaw, G., et al. 2009, ApJS, 180, 225.

\reference{} Houck, J.R., et al. 2004, ApJS, 154, 18.

\reference{} Howell, J.H., et al. 2009, ApJ, submitted.

\reference{} Ishida, C., 2004 PhD Thesis, University of Hawaii.

\reference{} Iwasawa, K., Sanders, D.B., Evans, A.S., Mazzarella, J.M., Armus, L., \& Surace, J. 2009a, ApJ, 695, 103L.

\reference{} Iwasawa, K., et al. 2009b, in preparation.

\reference{} Kim, D.-C., Veilleux, S., \& Sanders, D.B. 1998, ApJ, 508, 627.

\reference{} Kim, D.-C., Sanders, D.B., Veilleux, S., Mazzarella, J.M., \& Soifer, B.T. 1995, ApJS, 98, 129.

\reference{} Komossa, S., et al. 2003, ApJ, 582, 15L.

\reference{} Kong, X., Charlot, S., Brinchmann, J., \& Fall, S.M. 2004, MNRAS, 349, 769.

\reference{} Lacy, M. et al. 2004, ApJS, 154, 166.

\reference{} Lahuis, F., et al. 2007, ApJ, 659, 296.

\reference{} Le Floch, E., et al. 2005, ApJ, 632, 169L.

\reference{} Levenson, N.A., et al. 2007, ApJ, 654, L45.

%\reference{} Lonsdale, C.J., Smith, H.E., \& Lonsdale, C.J. 1995, ApJ, 438, 632.

\reference{} Lo, K.Y., Gao, Y., \& Gruendl, R.A. 1997, ApJ, 475, L103.

%\reference{} Lutz, D., Kunze, D., Spoon, H.W.W., \& Thornley, M.D. 1998, A\&A, 333, L75.

%\reference{} Lutz, D., et al. 2003, A\&A, 409, 867.

\reference{} Magnelli, B., et al. 2009, Astronomy \& Astrophysics, 496, 57.

\reference{} Martin, D.C., et al. 2007, ApJS, 173, 415.

\reference{} Mazzarella, J.M., et al. 2009, in preparation.

\reference{} Mazzarella, J.M. \& Balzano, V.A. 1986, ApJS, 62, 751.

\reference{} Mazzarella, J.M., Graham, J.R., Sanders, D.B., \& Djorgovski, S. 1993, ApJ, 409, 170.

\reference{} Mengel, S., Lehnert, M.D., Thatte, N., Tacconi-Garman, L., \& Genzel, R. 2001, ApJ, 550, 280.

\reference{} Mengel, S., Lehnert, M.D., Thatte, N., \& Genzel R. 2005, A \& A, 443, 41.

\reference{} Meurer, G.R., Heckman, T.M., \& Calzetti D. 1999, ApJ, 521, 64.

%\reference{} Mihos, C.J., \& Hernquist, L. 1996, ApJ, 464, 641.

\reference{} Mirabel, I.F., et al. 1998, Astronomy \& Astrophysics, 333, L1.

%\reference{} Moshir, et al. 1990, IRAS Faint Source Catalog, V2.0.

\reference{} Mould, J.R., et al. 2000, ApJ, 529, 786.

\reference{} Murphy, E.J., et al. 2006, ApJ, 651, L111.

\reference{} Murphy, T.W. Jr., Armus, L., Matthews, K., Soifer, B.T., Mazzarella, J.M., 
Shupe, D.L., Strauss, M.A., \& Neugebauer, G. 1996, AJ, 111, 1025.

\reference{} Murphy, T.W. Jr., Soifer, B.T., Matthews, K., Kiger, J.R., \& Armus, L. 1999, ApJ, 525, 85L.

\reference{} Murphy, T.W. Jr., Soifer, B.T., Matthews, K., \& Armus, L. 2001a, ApJ, 559, 201.

\reference{} Murphy, T.W. Jr., Soifer, B.T., Matthews, K., Armus, L., \& Kiger, J.R. 2001b, AJ, 121, 97.

\reference{} Petric, A., et al. 2009, in preparation.

\reference{} Ranalli, P., Comastri, A., \& Setti, G. 2003, A\&A, 399, 39.

\reference{} Read, A.M., Ponman, T.J. \& Strickland, D.K. 1997, MNRAS, 286, 626.

\reference{} Rhoads, J.E. 2000, PASP, 112, 703.

\reference{} Sanders, D.B., et al. 1988a, ApJ, 325, 74.
 
\reference{} Sanders, D.B., Soifer, B.T., Elias, J.H., Neugebauer G., \&
Matthews, K. 1988b, ApJ, 328, L35.

\reference{} Sanders, D.B., Scoville, N.Z., \& Soifer, B.T. 1991, ApJ, 370, 158.

\reference{} Sanders, D.B., \& Mirabel, I.F. 1996, ARA\&A, 34, 749.

\reference{} Sanders, D.B., Mazzarella, J.M., Kim, D.-C., Surace, J.A., \& Soifer, B.T. 2003, AJ, 126, 1607.

\reference{} Schecter, P. 1976, ApJ, 203, 297.

\reference{} Schmidt, M., \& Green, R.F. 1983, ApJ, 269, 352.

\reference {} Schiminovich, D., et al. 2007, ApJS, 173, 315.

\reference{} Scoville, N., et al. 2007, ApJS, 172, 1.

\reference{} Sirocky, M.M., Levenson, N.A., Elitzur, M., Spoon, H.W.W., \& Armus, L. 2008, ApJ, 678, 729.

\reference{} Skrutskie, M.F., et al. 2006, AJ, 131, 1163.

\reference{} Smith, J.D., et al. 2007, ApJ, 656, 770.

%\reference{} Solomon, P.M., Downes, D., Radford, S.J.E., \& Barrett, J.W. 1997, ApJ, 478, 144.

\reference{} Spoon, H.W.W.W., et al. 2004, ApJS, 154, 184.

\reference{} Spoon, H.W.W.W., et al. 2006, ApJ, 638, 759.

\reference{} Stern, D., et al. 2005 ApJ, 631, 163.

\reference{} Strickland, D.K., Heckman, T.M., Colbert, E.J.M., Hoopes, C.G., and Weaver. K.A. 2004a, ApJS, 151, 193.

\reference{} Strickland, D.K., Heckman, T.M., Colbert, E.J.M., Hoopes, C.G., and Weaver. K.A. 2004b, ApJ, 606, 829.

%\reference{} Sturm, E., et al. 2000, A \& A, 358, 481.

\reference{} Sturm, E., et al. 2002, A \& A, 393, 821.

\reference{} Tacconi, L.J., et al. 2002, ApJ, 580, 73.

%\reference{} Teng, S.H., et al. 2005, ApJ, 633, 664.

\reference{} Teng, S.H., et al. 2009, ApJ, in press.

\reference{} Tran, Q.D., et al. 2001, ApJ, 552, 527.

\reference{} van Dokkum, P.G. 2001, PASP, 113, 1420.

\reference{} Vavilkin, T., et al. 2009, in preparation.

\reference{} Veilleux, S., Kim, D.C., Sanders, D.B., Mazzarella, J.M. \& Soifer, B.T.
1995, ApJ Suppl., 98, 171.

\reference{} Veilleux, S., Sanders, D.B., \& Kim, D.-C. 1997, ApJ, 484, 92.

\reference{} Whitmore, B.C., \& Schweizer, F. 1995, AJ, 109, 960.

\reference{} Whitmore, B.C., et al. 2005, AJ, 130, 2104.

\reference{} Whitmore, B.C., Rupali, C., \& Fall, M.S. 2007, AJ, 133, 1067.

\reference{} Wright, G.S., James, P.A., Joseph, R.D., \& McLean, I.S. 1990, Nature, 344, 417.

\reference{} Xu, C.K., et al. 2007, ApJS, 173, 432.

\reference{} Yun, M.S., Reddy, N.A., \& Condon, J.J. 2001, ApJ, 554, 803.

\clearpage

\LongTables
%Table 1 -- The GOALS Sample
\begin{deluxetable*}{rlrrrrc}
\tabletypesize{\footnotesize}
%\tiny(5pt);\scriptsize(7pt);\footnotesize(8pt);\small(9pt);\normalsize(10pt)
\setlength{\tabcolsep}{0.04in} %Tighten up the columns. See AASTeX FAQ
%\rotate
\tablenum{1}
\tablewidth{0pt}
\tablecaption{The GOALS Sample}
\tablehead{
\colhead{\tt IRAS Name} &\colhead{\tt Optical ID} &\colhead{\tt RA} &\colhead{\tt Dec} &\colhead{\tt $V_{Helio}$} &\colhead{\tt $D_L$}   &\colhead{\tt $log(L_{ir}/L_{\odot})$}\\
\colhead{}              &\colhead{}               &\colhead{\tt J2000}    &\colhead{\tt J2000} &\colhead{\tt $km/s$}    &\colhead{\tt $Mpc$} &\colhead{}\\
\colhead{(1)}           &\colhead{(2)}            &\colhead{(3)}          &\colhead{(4)}       &\colhead{(5)}           &\colhead{(6)}       &\colhead{(7)}
}
\startdata
{\tt F00073+2538}&      {\tt NGC 0023}& {\tt 00h09m53.41s}&     {\tt +25d55m25.6s}&     {\tt 4566}&     {\tt 65.2}&     {\tt 11.12\phm{:}}      \\
{\tt F00085-1223}&      {\tt NGC 0034}& {\tt 00h11m06.55s}&    {\tt -12d06m26.3s}&    {\tt 5881}&     {\tt 84.1}&     {\tt 11.49\phm{:}}      \\
{\tt F00163-1039}&      {\tt Arp 256}&  {\tt 00h18m50.51s}&     {\tt -10d22m09.2s}&     {\tt 8159}&     {\tt 117.5}&    {\tt 11.48\phm{:}}      \\
{\tt F00344-3349}&      {\tt ESO 350-IG 038}&   {\tt 00h36m52.25s}&     {\tt -33d33m18.1s}&    {\tt 6175}&     {\tt 89.0}&     {\tt 11.28\phm{:}}      \\
{\tt F00402-2349}&      {\tt NGC 0232}& {\tt 00h42m45.82s}&     {\tt -23d33m40.9s}&     {\tt 6647}&     {\tt 95.2}&     {\tt 11.44\phm{:}}      \\
{\tt F00506+7248}&      {\tt MCG +12-02-001}&   {\tt 00h54m03.61s}&     {\tt +73d05m11.8s}&     {\tt 4706}&     {\tt 69.8}&     {\tt 11.50\phm{:}}      \\
{\tt F00548+4331}&      {\tt NGC 0317B}&        {\tt 00h57m40.45s}&    {\tt +43d47m32.1s}&    {\tt 5429}&     {\tt 77.8}&     {\tt 11.19\phm{:}}      \\
{\tt F01053-1746}&      {\tt IC 1623}&  {\tt 01h07m47.18s}&    {\tt -17d30m25.3s}&    {\tt 6016}&     {\tt 85.5}&     {\tt 11.71\phm{:}}      \\
{\tt F01076-1707}&      {\tt MCG -03-04-014}&   {\tt 01h10m08.96s}&     {\tt -16d51m09.8s}&     {\tt 10040}&    {\tt 144.0}&    {\tt 11.65\phm{:}}      \\
{\tt F01159-4443}&      {\tt ESO 244-G012}&     {\tt 01h18m08.15s}&     {\tt -44d27m51.2s}&     {\tt 6307}&     {\tt 91.5}&     {\tt 11.38:}    \\
{\tt F01173+1405}&      {\tt CGCG 436-030}&     {\tt 01h20m02.72s}&    {\tt +14d21m42.9s}&    {\tt 9362}&     {\tt 134.0}&    {\tt 11.69\phm{:}}      \\
{\tt F01325-3623}&      {\tt ESO 353-G020}&     {\tt 01h34m51.28s}&     {\tt -36d08m14.0s}&     {\tt 4797}&     {\tt 68.8}&     {\tt 11.06\phm{:}}      \\
{\tt F01341-3735}&      {\tt RR 032}&   {\tt 01h36m23.79s}&     {\tt -37d19m51.7s}&     {\tt 5191}&     {\tt 74.6}&     {\tt 11.16:}    \\
{\tt F01364-1042}&  & {\tt 01h38m52.92s}&    {\tt -10d27m11.4s}&    {\tt 14464}&    {\tt 210.0}&    {\tt 11.85:}    \\
{\tt F01417+1651}&      {\tt III Zw 035}&       {\tt 01h44m30.45s}&     {\tt +17d06m05.0s}&     {\tt 8375}&     {\tt 119.0}&    {\tt 11.64:}    \\
{\tt F01484+2220}&      {\tt NGC 0695}& {\tt 01h51m14.24s}&    {\tt +22d34m56.5s}&    {\tt 9735}&     {\tt 139.0}&    {\tt 11.68\phm{:}}      \\
{\tt F01519+3640}&      {\tt UGC 01385}&        {\tt 01h54m53.79s}&     {\tt +36d55m04.6s}&     {\tt 5621}&     {\tt 79.8}&     {\tt 11.05:}    \\
{\tt F02071-1023}&      {\tt NGC 0838}& {\tt 02h09m38.58s}&    {\tt -10d08m46.3s}&    {\tt 3851}&     {\tt 53.8}&     {\tt 11.05:}    \\
{\tt F02070+3857}&      {\tt NGC 0828}& {\tt 02h10m09.57s}&     {\tt +39d11m25.3s}&     {\tt 5374}&     {\tt 76.3}&     {\tt 11.36\phm{:}}      \\
{\tt F02114+0456}&      {\tt IC 0214}&  {\tt 02h14m05.59s}&     {\tt +05d10m23.7s}&     {\tt 9061}&     {\tt 129.0}&    {\tt 11.43\phm{:}}      \\
{\tt F02152+1418}&      {\tt NGC 0877}& {\tt 02h17m59.64s}&     {\tt +14d32m38.6s}&     {\tt 3913}&     {\tt 54.6}&     {\tt 11.10\phm{:}}      \\
{\tt F02203+3158}&      {\tt MCG +05-06-036}&   {\tt 02h23m21.99s}&    {\tt +32d11m49.5s}&    {\tt 10106}&    {\tt 145.0}&    {\tt 11.64\phm{:}}      \\
{\tt F02208+4744}&      {\tt UGC 01845}&        {\tt 02h24m07.98s}&    {\tt +47d58m11.0s}&    {\tt 4679}&     {\tt 67.0}&     {\tt 11.12\phm{:}}      \\
{\tt F02281-0309}&      {\tt NGC 0958}& {\tt 02h30m42.83s}&     {\tt -02d56m20.4s}&     {\tt 5738}&     {\tt 80.6}&     {\tt 11.20\phm{:}}      \\
{\tt F02345+2053}&      {\tt NGC 0992}& {\tt 02h37m25.49s}&     {\tt +21d06m03.0s}&     {\tt 4141}&     {\tt 58.0}&     {\tt 11.07\phm{:}}      \\
{\tt F02401-0013}&      {\tt NGC 1068}& {\tt 02h42m40.71s}&    {\tt -00d00m47.8s}&    {\tt 1137}&     {\tt 15.9}&     {\tt 11.40\phm{:}}      \\
{\tt F02435+1253}&      {\tt UGC 02238}&        {\tt 02h46m17.49s}&     {\tt +13d05m44.4s}&     {\tt 6560}&     {\tt 92.4}&     {\tt 11.33:}    \\
{\tt F02437+2122}& & {\tt 02h46m39.15s}&    {\tt +21d35m10.3s}&    {\tt 6987}&     {\tt 98.8}&     {\tt 11.16:}    \\
{\tt F02512+1446}&      {\tt UGC 02369}&        {\tt 02h54m01.78s}&     {\tt +14d58m24.9s}&     {\tt 9558}&     {\tt 136.0}&    {\tt 11.67\phm{:}}      \\
{\tt F03117+4151}&      {\tt UGC 02608}&        {\tt 03h15m01.42s}&    {\tt +42d02m09.4s}&    {\tt 6998}&     {\tt 100.0}&    {\tt 11.41\phm{:}}      \\
{\tt F03164+4119}&      {\tt NGC 1275}& {\tt 03h19m48.16s}&   {\tt +41d30m42.1s}&   {\tt 5264}&     {\tt 75.0}&     {\tt 11.26\phm{:}}      \\
{\tt F03217+4022}& & {\tt 03h25m05.38s}&    {\tt +40d33m29.0s}&    {\tt 7007}&     {\tt 100.0}&    {\tt 11.33\phm{:}}      \\
{\tt F03316-3618}&      {\tt NGC 1365}& {\tt 03h33m36.37s}&    {\tt -36d08m25.4s}&    {\tt 1636}&     {\tt 17.9}&     {\tt 11.00\phm{:}}      \\
{\tt F03359+1523}& & {\tt 03h38m46.70s}&     {\tt +15d32m55.0s}&     {\tt 10613}&    {\tt 152.0}&    {\tt 11.55:}    \\
{\tt F03514+1546}&      {\tt CGCG 465-012}&     {\tt 03h54m16.08s}&    {\tt +15d55m43.4s}&    {\tt 6662}&     {\tt 94.3}&     {\tt 11.20:}    \\
{\tt 03582+6012 }& &  {\tt 04h02m32.48s}&     {\tt +60d20m40.1s}&     {\tt 8997}&     {\tt 131.0}&    {\tt 11.43:}    \\
{\tt F04097+0525}&      {\tt UGC 02982}&        {\tt 04h12m22.45s}&     {\tt +05d32m50.6s}&     {\tt 5305}&     {\tt 74.9}&     {\tt 11.20\phm{:}}      \\
{\tt F04118-3207}&      {\tt ESO 420-G013}&     {\tt 04h13m49.69s}&     {\tt -32d00m25.1s}&     {\tt 3570}&     {\tt 51.0}&     {\tt 11.07\phm{:}}      \\
{\tt F04191-1855}&      {\tt ESO 550-IG 025}&   {\tt 04h21m20.02s}&     {\tt -18d48m47.6s}&     {\tt 9621}&     {\tt 138.5}&    {\tt 11.51\phm{:}}      \\
{\tt F04210-4042}&      {\tt NGC 1572}& {\tt 04h22m42.81s}&     {\tt -40d36m03.3s}&     {\tt 6111}&     {\tt 88.6}&     {\tt 11.30\phm{:}}      \\
{\tt 04271+3849 }& &  {\tt 04h30m33.09s}&     {\tt +38d55m47.7s}&     {\tt 5640}&     {\tt 80.8}&     {\tt 11.11:}    \\
{\tt F04315-0840}&      {\tt NGC 1614}& {\tt 04h33m59.85s}&    {\tt -08d34m44.0s}&    {\tt 4778}&     {\tt 67.8}&     {\tt 11.65\phm{:}}      \\
{\tt F04326+1904}&      {\tt UGC 03094}&        {\tt 04h35m33.83s}&     {\tt +19d10m18.2s}&     {\tt 7408}&     {\tt 106.0}&    {\tt 11.41\phm{:}}      \\
{\tt F04454-4838}&      {\tt ESO 203-IG001}&    {\tt 04h46m49.50s}&     {\tt -48d33m32.9s}&     {\tt 15862}&    {\tt 235.0}&    {\tt 11.86:}    \\
{\tt F04502-3304}&      {\tt MCG -05-12-006}&   {\tt 04h52m04.96s}&     {\tt -32d59m25.6s}&     {\tt 5622}&     {\tt 81.3}&     {\tt 11.17\phm{:}}      \\
{\tt F05053-0805}&      {\tt NGC 1797}& {\tt 05h07m44.88s}&     {\tt -08d01m08.7s}&     {\tt 4441}&     {\tt 63.4}&     {\tt 11.04\phm{:}}      \\
{\tt F05054+1718}&      {\tt CGCG 468-002}&     {\tt 05h08m20.5s}&      {\tt +17d21m58s}&       {\tt 5454}&     {\tt 77.9}&     {\tt 11.22\phm{:}}      \\
{\tt 05083+2441 }& &  {\tt 05h11m25.88s}&     {\tt +24d45m18.3s}&     {\tt 6915}&     {\tt 99.2}&     {\tt 11.26:}    \\
{\tt F05081+7936}&      {\tt VII Zw 031}&       {\tt 05h16m46.44s}&     {\tt +79d40m12.6s}&     {\tt 16090}&    {\tt 240.0}&    {\tt 11.99\phm{:}}      \\
{\tt 05129+5128 }& &  {\tt 05h16m56.10s}&     {\tt +51d31m56.5s}&     {\tt 8224}&     {\tt 120.0}&    {\tt 11.42\phm{:}}      \\
{\tt F05189-2524}& & {\tt 05h21m01.47s}&    {\tt -25d21m45.4s}&    {\tt 12760}&    {\tt 187.0}&    {\tt 12.16\phm{:}}      \\
{\tt F05187-1017}& & {\tt 05h21m06.54s}&    {\tt -10d14m46.7s}&    {\tt 8474}&     {\tt 122.0}&    {\tt 11.30:}    \\
{\tt 05223+1908 }& &  {\tt 05h25m16.50s}&     {\tt +19d10m46.0s}&     {\tt 8867}&     {\tt 128.0}&    {\tt 11.65:}    \\
{\tt 05368+4940 }&      {\tt MCG +08-11-002}&   {\tt 05h40m43.71s}&     {\tt +49d41m41.5s}&     {\tt 5743}&     {\tt 83.7}&     {\tt 11.46\phm{:}}      \\
{\tt F05365+6921}&      {\tt NGC 1961}& {\tt 05h42m04.65s}&   {\tt +69d22m42.4s}&   {\tt 3934}&     {\tt 59.0}&     {\tt 11.06\phm{:}}      \\
{\tt F05414+5840}&      {\tt UGC 03351}&        {\tt 05h45m47.88s}&     {\tt +58d42m03.9s}&     {\tt 4455}&     {\tt 65.8}&     {\tt 11.28\phm{:}}      \\
{\tt 05442+1732 }& &  {\tt 05h47m11.18s}&     {\tt +17d33m46.7s}&     {\tt 5582}&     {\tt 80.5}&     {\tt 11.30\phm{:}}      \\
{\tt F06076-2139}& & {\tt 06h09m45.81s}&     {\tt -21d40m23.7s}&     {\tt 11226}&    {\tt 165.0}&    {\tt 11.65\phm{:}}      \\
{\tt F06052+8027}&      {\tt UGC 03410}&        {\tt 06h14m29.63s}&     {\tt +80d26m59.6s}&     {\tt 3921}&     {\tt 59.7}&     {\tt 11.10:}    \\
{\tt F06107+7822}&      {\tt NGC 2146}& {\tt 06h18m37.71s}&     {\tt +78d21m25.3s}&     {\tt 893}&      {\tt 17.5}&     {\tt 11.12\phm{:}}      \\
{\tt F06259-4708}&      {\tt ESO 255-IG007}&    {\tt 06h27m22.45s}&     {\tt -47d10m48.7s}&     {\tt 11629}&    {\tt 173.0}&    {\tt 11.90\phm{:}}      \\
{\tt F06295-1735}&      {\tt ESO 557-G002}&     {\tt 06h31m47.22s}&     {\tt -17d37m17.3s}&     {\tt 6385}&     {\tt 93.6}&     {\tt 11.25\phm{:}}      \\
{\tt F06538+4628}&      {\tt UGC 3608}& {\tt 06h57m34.45s}&     {\tt +46d24m10.8s}&     {\tt 6401}&     {\tt 94.3}&     {\tt 11.34\phm{:}}      \\
{\tt F06592-6313}& & {\tt 06h59m40.25s}&     {\tt -63d17m52.9s}&     {\tt 6882}&     {\tt 104.0}&    {\tt 11.24\phm{:}}      \\
{\tt F07027-6011}&      {\tt AM 0702-601 }&     {\tt 07h03m26.37s}&    {\tt -60d16m03.7s}&    {\tt 9390}&     {\tt 141.0}&    {\tt 11.64\phm{:}}      \\
{\tt 07063+2043 }&      {\tt NGC 2342}& {\tt 07h09m18.08s}&     {\tt +20d38m09.5s}&     {\tt 5276}&     {\tt 78.0}&     {\tt 11.31:}    \\
{\tt F07160-6215}&      {\tt NGC 2369}& {\tt 07h16m37.73s}&     {\tt -62d20m37.4s}&     {\tt 3240}&     {\tt 47.6}&     {\tt 11.16\phm{:}}      \\
{\tt 07251-0248 }& &  {\tt 07h27m37.55s}&     {\tt -02d54m54.1s}&     {\tt 26249}&    {\tt 400.0}&    {\tt 12.39:}    \\
{\tt F07256+3355}&      {\tt NGC 2388}& {\tt 07h28m53.44s}&    {\tt +33d49m08.7s}&    {\tt 4134}&     {\tt 62.1}&     {\tt 11.28:}    \\
{\tt F07329+1149}&      {\tt MCG +02-20-003}&   {\tt 07h35m43.37s}&     {\tt +11d42m33.5s}&     {\tt 4873}&     {\tt 72.8}&     {\tt 11.13:}    \\
{\tt 08355-4944 }& &  {\tt 08h37m01.82s}&     {\tt -49d54m30.2s}&     {\tt 7764}&     {\tt 118.0}&    {\tt 11.62\phm{:}}      \\
{\tt F08339+6517}& & {\tt 08h38m23.18s}&    {\tt +65d07m15.2s}&    {\tt 5730}&     {\tt 86.3}&     {\tt 11.11\phm{:}}      \\
{\tt F08354+2555}&      {\tt NGC 2623}& {\tt 08h38m24.08s}&    {\tt +25d45m16.6s}&    {\tt 5549}&     {\tt 84.1}&     {\tt 11.60\phm{:}}      \\
{\tt 08424-3130 }&      {\tt ESO 432-IG006}&    {\tt 08h44m28.07s}&     {\tt -31d41m40.6s}&     {\tt 4846}&     {\tt 74.4}&     {\tt 11.08\phm{:}}      \\
{\tt F08520-6850}&      {\tt ESO 060-IG 016}&   {\tt 08h52m31.29s}&     {\tt -69d01m57.0s}&     {\tt 13885}&    {\tt 210.0}&    {\tt 11.82:}    \\
{\tt F08572+3915}& & {\tt 09h00m25.39s}&    {\tt +39d03m54.4s}&    {\tt 17493}&    {\tt 264.0}&    {\tt 12.16\phm{:}}      \\
{\tt 09022-3615 }& &  {\tt 09h04m12.70s}&     {\tt -36d27m01.1s}&     {\tt 17880}&    {\tt 271.0}&    {\tt 12.31\phm{:}}      \\
{\tt F09111-1007}& & {\tt 09h13m37.61s}&     {\tt -10d19m24.8s}&     {\tt 16231}&    {\tt 246.0}&    {\tt 12.06\phm{:}}      \\
{\tt F09126+4432}&      {\tt UGC 04881}&        {\tt 09h15m55.11s}&     {\tt +44d19m54.1s}&     {\tt 11851}&    {\tt 178.0}&    {\tt 11.74\phm{:}}      \\
{\tt F09320+6134}&      {\tt UGC 05101}&        {\tt 09h35m51.65s}&    {\tt +61d21m11.3s}&    {\tt 11802}&    {\tt 177.0}&    {\tt 12.01:}    \\
{\tt F09333+4841}&      {\tt MCG +08-18-013}&   {\tt 09h36m37.19s}&     {\tt +48d28m27.7s}&     {\tt 7777}&     {\tt 117.0}&    {\tt 11.34\phm{:}}      \\
{\tt F09437+0317}&      {\tt Arp 303 }& {\tt 09h46m20.71s}&     {\tt +03d03m30.5s}&     {\tt 5996}&     {\tt 92.9}&     {\tt 11.23:}    \\
{\tt F10015-0614}&      {\tt NGC 3110}& {\tt 10h04m02.11s}&     {\tt -06d28m29.2s}&     {\tt 5054}&     {\tt 79.5}&     {\tt 11.37:}    \\
{\tt F10038-3338}&      {\tt ESO 374-IG 032\tablenotemark{a}}&  {\tt 10h06m04.8s}&     {\tt -33d53m15.0s}&     {\tt 10223}&    {\tt 156.0}&    {\tt 11.78\phm{:}}      \\
{\tt F10173+0828}& & {\tt 10h20m00.21s}&    {\tt +08d13m33.8s}&    {\tt 14716}&    {\tt 224.0}&    {\tt 11.86\phm{:}}      \\
{\tt F10196+2149}&      {\tt NGC 3221}& {\tt 10h22m19.98s}&     {\tt +21d34m10.5s}&     {\tt 4110}&     {\tt 65.7}&     {\tt 11.09\phm{:}}      \\
{\tt F10257-4339}&      {\tt NGC 3256}& {\tt 10h27m51.27s}&     {\tt -43d54m13.8s}&     {\tt 2804}&     {\tt 38.9}&     {\tt 11.64\phm{:}}      \\
{\tt F10409-4556}&      {\tt ESO 264-G036}&     {\tt 10h43m07.67s}&     {\tt -46d12m44.6s}&     {\tt 6299}&     {\tt 100.0}&    {\tt 11.32\phm{:}}      \\
{\tt F10567-4310}&      {\tt ESO 264-G057}&     {\tt 10h59m01.79s}&     {\tt -43d26m25.7s}&     {\tt 5156}&     {\tt 83.3}&     {\tt 11.14:}    \\
{\tt F10565+2448}& & {\tt 10h59m18.14s}&     {\tt +24d32m34.3s}&     {\tt 12921}&    {\tt 197.0}&    {\tt 12.08\phm{:}}      \\
{\tt F11011+4107}&      {\tt MCG +07-23-019}&   {\tt 11h03m53.20s}&     {\tt +40d50m57.0s}&     {\tt 10350}&    {\tt 158.0}&    {\tt 11.62\phm{:}}      \\
{\tt F11186-0242}&      {\tt CGCG 011-076}&     {\tt 11h21m12.26s}&    {\tt -02d59m03.5s}&    {\tt 7464}&     {\tt 117.0}&    {\tt 11.43\phm{:}}      \\
{\tt F11231+1456}& & {\tt 11h25m47.30s}&     {\tt +14d40m21.1s}&    {\tt 10192}&    {\tt 157.0}&    {\tt 11.64\phm{:}}      \\
{\tt F11255-4120}&      {\tt ESO 319-G022}&     {\tt 11h27m54.08s}&     {\tt -41d36m52.4s}&     {\tt 4902}&     {\tt 80.0}&     {\tt 11.12\phm{:}}      \\
{\tt F11257+5850}&      {\tt NGC 3690}& {\tt 11h28m32.25s}&     {\tt +58d33m44.0s}&     {\tt 3093}&     {\tt 50.7}&     {\tt 11.93\phm{:}}      \\
{\tt F11506-3851}&      {\tt ESO 320-G030}&     {\tt 11h53m11.72s}&     {\tt -39d07m48.9s}&     {\tt 3232}&     {\tt 41.2}&     {\tt 11.17\phm{:}}      \\
{\tt F12043-3140}&      {\tt ESO 440-IG058}&    {\tt 12h06m51.82s}&     {\tt -31d56m53.1s}&     {\tt 6956}&     {\tt 112.0}&    {\tt 11.43\phm{:}}      \\
{\tt F12112+0305}& & {\tt 12h13m46.00s}&     {\tt +02d48m38.0s}&     {\tt 21980}&    {\tt 340.0}&    {\tt 12.36:}    \\
{\tt F12116+5448}&      {\tt NGC 4194}& {\tt 12h14m09.47s}&    {\tt +54d31m36.6s}&    {\tt 2501}&     {\tt 43.0}&     {\tt 11.10\phm{:}}      \\
{\tt F12115-4656}&      {\tt ESO 267-G030}&     {\tt 12h14m12.84s}&     {\tt -47d13m43.2s}&     {\tt 5543}&     {\tt 97.1}&     {\tt 11.25:}    \\
{\tt 12116-5615 }& &  {\tt 12h14m22.10s}&     {\tt -56d32m33.2s}&     {\tt 8125}&     {\tt 128.0}&    {\tt 11.65\phm{:}}      \\
{\tt F12224-0624}& & {\tt 12h25m03.91s}&     {\tt -06d40m52.6s}&     {\tt 7902}&     {\tt 125.0}&    {\tt 11.36:}    \\
{\tt F12243-0036}&      {\tt NGC 4418}& {\tt 12h26m54.62s}&    {\tt -00d52m39.2s}&    {\tt 2179}&     {\tt 36.5}&     {\tt 11.19\phm{:}}      \\
{\tt F12540+5708}&      {\tt UGC 08058    }&    {\tt 12h56m14.24s}&   {\tt +56d52m25.2s}&   {\tt 12642}&    {\tt 192.0}&    {\tt 12.57\phm{:}}      \\
{\tt F12590+2934}&      {\tt NGC 4922}& {\tt 13h01m24.89s}&     {\tt +29d18m40.0s}&     {\tt 7071}&     {\tt 111.0}&    {\tt 11.38\phm{:}}      \\
{\tt F12592+0436}&      {\tt CGCG 043-099}&     {\tt 13h01m50.80s}&     {\tt +04d20m00.0s}&     {\tt 11237}&    {\tt 175.0}&    {\tt 11.68\phm{:}}      \\
{\tt F12596-1529}&      {\tt MCG -02-33-098}&   {\tt 13h02m19.70s}&     {\tt -15d46m03.0s}&     {\tt 4773}&     {\tt 78.7}&     {\tt 11.17\phm{:}}      \\
{\tt F13001-2339}&      {\tt ESO 507-G070}&     {\tt 13h02m52.35s}&     {\tt -23d55m17.7s}&     {\tt 6506}&     {\tt 106.0}&    {\tt 11.56\phm{:}}      \\
{\tt 13052-5711 }& &  {\tt 13h08m18.73s}&     {\tt -57d27m30.2s}&     {\tt 6364}&     {\tt 106.0}&    {\tt 11.40:}    \\
%{\tt F13097-1531}&      {\tt NGC 5010\tablenotemark{b}}&        {\tt 13h12m26.3s}&     {\tt -15d47m52.3s}&     {\tt 2975}&     {\tt 44.8}&     {\tt 10.84\phm{:}}      \\
{\tt F13126+2453}&      {\tt IC 0860}&  {\tt 13h15m03.53s}&    {\tt +24d37m07.9s}&    {\tt 3347}&     {\tt 56.8}&     {\tt 11.14:}    \\
{\tt 13120-5453 }& &  {\tt 13h15m06.35s}&     {\tt -55d09m22.7s}&     {\tt 9222}&     {\tt 144.0}&    {\tt 12.32\phm{:}}      \\
{\tt F13136+6223}&      {\tt VV 250a}&  {\tt 13h15m35.06s}&    {\tt +62d07m28.6s}&    {\tt 9313}&     {\tt 142.0}&    {\tt 11.81\phm{:}}      \\
{\tt F13182+3424}&      {\tt UGC 08387}&        {\tt 13h20m35.34s}&    {\tt +34d08m22.2s}&    {\tt 6985}&     {\tt 110.0}&    {\tt 11.73\phm{:}}      \\
{\tt F13188+0036}&      {\tt NGC 5104}& {\tt 13h21m23.08s}&    {\tt +00d20m32.7s}&    {\tt 5578}&     {\tt 90.8}&     {\tt 11.27\phm{:}}      \\
{\tt F13197-1627}&      {\tt MCG -03-34-064}&   {\tt 13h22m24.46s}&    {\tt -16d43m42.9s}&    {\tt 4959}&     {\tt 82.2}&     {\tt 11.28\phm{:}}      \\
{\tt F13229-2934}&      {\tt NGC 5135}& {\tt 13h25m44.06s}&     {\tt -29d50m01.2s}&     {\tt 4105}&     {\tt 60.9}&     {\tt 11.30\phm{:}}      \\
{\tt 13242-5713 }&      {\tt ESO 173-G015}&     {\tt 13h27m23.78s}&     {\tt -57d29m22.2s}&     {\tt 2918}&     {\tt 34.0}&     {\tt 11.38:}    \\
{\tt F13301-2356}&      {\tt IC 4280}&  {\tt 13h32m53.40s}&     {\tt -24d12m25.7s}&     {\tt 4889}&     {\tt 82.4}&     {\tt 11.15\phm{:}}      \\
{\tt F13362+4831}&      {\tt NGC 5256}& {\tt 13h38m17.52s}&     {\tt +48d16m36.7s}&     {\tt 8341}&     {\tt 129.0}&    {\tt 11.56\phm{:}}      \\
{\tt F13373+0105}&      {\tt Arp 240}&  {\tt 13h39m55.00s}&     {\tt +00d50m07.0s}&     {\tt 6778}&     {\tt 108.5}&    {\tt 11.62\phm{:}}      \\
{\tt F13428+5608}&      {\tt UGC 08696}&        {\tt 13h44m42.11s}&    {\tt +55d53m12.6s}&    {\tt 11326}&    {\tt 173.0}&    {\tt 12.21\phm{:}}      \\
{\tt F13470+3530}&      {\tt UGC 08739}&        {\tt 13h49m13.93s}&     {\tt +35d15m26.8s}&     {\tt 5032}&     {\tt 81.4}&     {\tt 11.15:}    \\
{\tt F13478-4848}&      {\tt ESO 221-IG010}&    {\tt 13h50m56.94s}&     {\tt -49d03m19.5s}&     {\tt 3099}&     {\tt 62.9}&     {\tt 11.22\phm{:}}      \\
{\tt F13497+0220}&      {\tt NGC 5331}& {\tt 13h52m16.29s}&     {\tt +02d06m17.0s}&     {\tt 9906}&     {\tt 155.0}&    {\tt 11.66\phm{:}}      \\
{\tt F13564+3741}&      {\tt Arp 84}&   {\tt 13h58m35.81s}&    {\tt +37d26m20.3s}&    {\tt 3482}&     {\tt 58.7}&     {\tt 11.08\phm{:}}      \\
{\tt F14179+4927}&      {\tt CGCG 247-020}&     {\tt 14h19m43.25s}&    {\tt +49d14m11.7s}&    {\tt 7716}&     {\tt 120.0}&    {\tt 11.39\phm{:}}      \\
{\tt F14280+3126}&      {\tt NGC 5653}& {\tt 14h30m10.42s}&     {\tt +31d12m55.8s}&     {\tt 3562}&     {\tt 60.2}&     {\tt 11.13\phm{:}}      \\
{\tt F14348-1447}& & {\tt 14h37m38.37s}&    {\tt -15d00m22.8s}&    {\tt 24802}&    {\tt 387.0}&    {\tt 12.39:}    \\
{\tt F14378-3651}& & {\tt 14h40m59.01s}&     {\tt -37d04m32.0s}&     {\tt 20277}&    {\tt 315.0}&    {\tt 12.23:}    \\
{\tt F14423-2039}&      {\tt NGC 5734}& {\tt 14h45m09.05s}&     {\tt -20d52m13.7s}&     {\tt 4121}&     {\tt 67.1}&     {\tt 11.15:}    \\
{\tt F14547+2449}&      {\tt VV 340a}&  {\tt 14h57m00.68s}&     {\tt +24d37m02.7s}&     {\tt 10094}&    {\tt 157.0}&    {\tt 11.74\phm{:}}      \\
{\tt F14544-4255}&      {\tt IC 4518}&  {\tt 14h57m42.82s}&     {\tt -43d07m54.3s}&     {\tt 4763}&     {\tt 80.0}&     {\tt 11.23\phm{:}}      \\
{\tt F15107+0724}&      {\tt CGCG 049-057}&     {\tt 15h13m13.09s}&    {\tt +07d13m31.8s}&    {\tt 3897}&     {\tt 65.4}&     {\tt 11.35:}    \\
{\tt F15163+4255}&      {\tt VV 705}&   {\tt 15h18m06.28s}&     {\tt +42d44m41.2s}&     {\tt 11944}&    {\tt 183.0}&    {\tt 11.92\phm{:}}      \\
{\tt 15206-6256 }&      {\tt ESO 099-G004}&     {\tt 15h24m58.19s}&     {\tt -63d07m34.2s}&     {\tt 8779}&     {\tt 137.0}&    {\tt 11.74\phm{:}}      \\
{\tt F15250+3608}& & {\tt 15h26m59.40s}&    {\tt +35d58m37.5s}&    {\tt 16535}&    {\tt 254.0}&    {\tt 12.08\phm{:}}      \\
{\tt F15276+1309}&      {\tt NGC 5936}& {\tt 15h30m00.84s}&     {\tt +12d59m21.5s}&     {\tt 4004}&     {\tt 67.1}&     {\tt 11.14\phm{:}}      \\
{\tt F15327+2340}&      {\tt UGC 09913}&        {\tt 15h34m57.12s}&    {\tt +23d30m11.5s}&    {\tt 5434}&     {\tt 87.9}&     {\tt 12.28\phm{:}}      \\
{\tt F15437+0234}&      {\tt NGC 5990}& {\tt 15h46m16.37s}&     {\tt +02d24m55.7s}&     {\tt 3839}&     {\tt 64.4}&     {\tt 11.13\phm{:}}      \\
{\tt F16030+2040}&      {\tt NGC 6052}& {\tt 16h05m13.05s}&     {\tt +20d32m32.6s}&     {\tt 4739}&     {\tt 77.6}&     {\tt 11.09\phm{:}}      \\
{\tt F16104+5235}&      {\tt NGC 6090}& {\tt 16h11m40.70s}&     {\tt +52d27m24.0s}&     {\tt 8947}&     {\tt 137.0}&    {\tt 11.58\phm{:}}      \\
{\tt F16164-0746}& & {\tt 16h19m11.79s}&     {\tt -07d54m02.8s}&     {\tt 8140}&     {\tt 128.0}&    {\tt 11.62:}    \\
{\tt F16284+0411}&      {\tt CGCG 052-037}&     {\tt 16h30m56.54s}&     {\tt +04d04m58.4s}&     {\tt 7342}&     {\tt 116.0}&    {\tt 11.45\phm{:}}      \\
{\tt 16304-6030 }&      {\tt NGC 6156}& {\tt 16h34m52.55s}&     {\tt -60d37m07.7s}&     {\tt 3263}&     {\tt 48.0}&     {\tt 11.14\phm{:}}      \\
{\tt F16330-6820}&      {\tt ESO 069-IG006}&    {\tt 16h38m12.65s}&     {\tt -68d26m42.6s}&     {\tt 13922}&    {\tt 212.0}&    {\tt 11.98\phm{:}}      \\
{\tt F16399-0937}& & {\tt 16h42m40.21s}&     {\tt -09d43m14.4s}&     {\tt 8098}&     {\tt 128.0}&    {\tt 11.63:}    \\
{\tt F16443-2915}&      {\tt ESO 453-G005}&     {\tt 16h47m31.06s}&     {\tt -29d21m21.6s}&     {\tt 6260}&     {\tt 100.0}&    {\tt 11.37:}    \\
{\tt F16504+0228}&      {\tt NGC 6240}& {\tt 16h52m58.89s}&     {\tt +02d24m03.4s}&     {\tt 7339}&     {\tt 116.0}&    {\tt 11.93\phm{:}}      \\
{\tt F16516-0948}& & {\tt 16h54m24.03s}&     {\tt -09d53m20.9s}&     {\tt 6755}&     {\tt 107.0}&    {\tt 11.31:}    \\
{\tt F16577+5900}&      {\tt NGC 6286}& {\tt 16h58m31.38s}&     {\tt +58d56m10.5s}&     {\tt 5501}&     {\tt 85.7}&     {\tt 11.37\phm{:}}      \\
{\tt F17132+5313}& & {\tt 17h14m20.00s}&     {\tt +53d10m30.0s}&     {\tt 15270}&    {\tt 232.0}&    {\tt 11.96\phm{:}}      \\
{\tt F17138-1017}& & {\tt 17h16m35.79s}&     {\tt -10d20m39.4s}&     {\tt 5197}&     {\tt 84.0}&     {\tt 11.49\phm{:}}      \\
{\tt F17207-0014}& & {\tt 17h23m21.95s}&    {\tt -00d17m00.9s}&    {\tt 12834}&    {\tt 198.0}&    {\tt 12.46:}    \\
{\tt F17222-5953}&      {\tt ESO 138-G027}&     {\tt 17h26m43.34s}&     {\tt -59d55m55.3s}&     {\tt 6230}&     {\tt 98.3}&     {\tt 11.41\phm{:}}      \\
{\tt F17530+3447}&      {\tt UGC 11041}&        {\tt 17h54m51.82s}&    {\tt +34d46m34.4s}&    {\tt 4881}&     {\tt 77.5}&     {\tt 11.11\phm{:}}      \\
{\tt F17548+2401}&      {\tt CGCG 141-034}&     {\tt 17h56m56.63s}&    {\tt +24d01m01.6s}&    {\tt 5944}&     {\tt 93.4}&     {\tt 11.20\phm{:}}      \\
{\tt 17578-0400 }& &  {\tt 18h00m31.90s}&     {\tt -04d00m53.3s}&     {\tt 4210}&     {\tt 68.5}&     {\tt 11.48:}    \\
{\tt 18090+0130 }& &  {\tt 18h11m35.91s}&     {\tt +01d31m41.3s}&     {\tt 8662}&     {\tt 134.0}&    {\tt 11.65:}    \\
{\tt F18131+6820}&      {\tt NGC 6621}& {\tt 18h12m55.31s}&     {\tt +68d21m48.4s}&     {\tt 6191}&     {\tt 94.3}&     {\tt 11.29\phm{:}}      \\
{\tt F18093-5744}&      {\tt IC 4687}&  {\tt 18h13m39.63s}&    {\tt -57d43m31.3s}&    {\tt 5200}&     {\tt 81.9}&     {\tt 11.62:}    \\
{\tt F18145+2205}&      {\tt CGCG 142-034}&     {\tt 18h16m40.66s}&     {\tt +22d06m46.1s}&     {\tt 5599}&     {\tt 88.1}&     {\tt 11.18\phm{:}}      \\
{\tt F18293-3413}& & {\tt 18h32m41.13s}&    {\tt -34d11m27.5s}&    {\tt 5449}&     {\tt 86.0}&     {\tt 11.88\phm{:}}      \\
{\tt F18329+5950}&      {\tt NGC 6670}& {\tt 18h33m35.91s}&     {\tt +59d53m20.2s}&     {\tt 8574}&     {\tt 129.5}&    {\tt 11.65\phm{:}}      \\
{\tt F18341-5732}&      {\tt IC 4734}&  {\tt 18h38m25.70s}&     {\tt -57d29m25.6s}&     {\tt 4680}&     {\tt 73.4}&     {\tt 11.35\phm{:}}      \\
{\tt F18425+6036}&      {\tt NGC 6701}& {\tt 18h43m12.46s}&     {\tt +60d39m12.0s}&     {\tt 3965}&     {\tt 62.4}&     {\tt 11.12\phm{:}}      \\
{\tt F19120+7320}&      {\tt VV 414}&   {\tt 19h10m59.20s}&     {\tt +73d25m06.3s}&     {\tt 7528}&     {\tt 113.0}&    {\tt 11.49\phm{:}}      \\
{\tt F19115-2124}&      {\tt ESO 593-IG008}&    {\tt 19h14m30.90s}&     {\tt -21d19m07.0s}&     {\tt 14608}&    {\tt 222.0}&    {\tt 11.93\phm{:}}      \\
{\tt F19297-0406}& & {\tt 19h32m21.25s}&     {\tt -03d59m56.3s}&     {\tt 25701}&    {\tt 395.0}&    {\tt 12.45:}    \\
{\tt 19542+1110 }& &  {\tt 19h56m35.44s}&     {\tt +11d19m02.6s}&     {\tt 19473}&    {\tt 295.0}&    {\tt 12.12:}    \\
{\tt F19542-3804}&      {\tt ESO 339-G011}&     {\tt 19h57m37.54s}&     {\tt -37d56m08.4s}&     {\tt 5756}&     {\tt 88.6}&     {\tt 11.20\phm{:}}      \\
{\tt F20221-2458}&      {\tt NGC 6907}& {\tt 20h25m06.65s}&     {\tt -24d48m33.5s}&     {\tt 3190}&     {\tt 50.1}&     {\tt 11.11\phm{:}}      \\
{\tt 20264+2533 }&      {\tt MCG +04-48-002}&   {\tt 20h28m35.06s}&     {\tt +25d44m00.0s}&     {\tt 4167}&     {\tt 64.2}&     {\tt 11.11\phm{:}}      \\
{\tt F20304-0211}&      {\tt NGC 6926}& {\tt 20h33m06.11s}&     {\tt -02d01m39.0s}&     {\tt 5880}&     {\tt 89.1}&     {\tt 11.32\phm{:}}      \\
{\tt 20351+2521 }& &  {\tt 20h37m17.72s}&     {\tt +25d31m37.7s}&     {\tt 10102}&    {\tt 151.0}&    {\tt 11.61\phm{:}}      \\
{\tt F20550+1655}&      {\tt CGCG 448-020}&     {\tt 20h57m23.90s}&     {\tt +17d07m39.0s}&     {\tt 10822}&    {\tt 161.0}&    {\tt 11.94\phm{:}}      \\
{\tt F20551-4250}&      {\tt ESO 286-IG019}&    {\tt 20h58m26.79s}&     {\tt -42d39m00.3s}&     {\tt 12890}&    {\tt 193.0}&    {\tt 12.06\phm{:}}      \\
{\tt F21008-4347}&      {\tt ESO 286-G035}&     {\tt 21h04m11.18s}&     {\tt -43d35m33.0s}&     {\tt 5205}&     {\tt 79.1}&     {\tt 11.20:}    \\
{\tt 21101+5810 }& &  {\tt 21h11m30.40s}&     {\tt +58d23m03.2s}&     {\tt 11705}&    {\tt 174.0}&    {\tt 11.81:}    \\
{\tt F21330-3846}&      {\tt ESO 343-IG013}&    {\tt 21h36m10.83s}&     {\tt -38d32m37.9s}&     {\tt 5714}&     {\tt 85.8}&     {\tt 11.14\phm{:}}      \\
{\tt F21453-3511}&      {\tt NGC 7130}& {\tt 21h48m19.50s}&     {\tt -34d57m04.7s}&     {\tt 4842}&     {\tt 72.7}&     {\tt 11.42\phm{:}}      \\
{\tt F22118-2742}&      {\tt ESO 467-G027}&     {\tt 22h14m39.92s}&     {\tt -27d27m50.3s}&     {\tt 5217}&     {\tt 77.3}&     {\tt 11.08\phm{:}}      \\
{\tt F22132-3705}&      {\tt IC 5179}&  {\tt 22h16m09.10s}&     {\tt -36d50m37.4s}&     {\tt 3422}&     {\tt 51.4}&     {\tt 11.24\phm{:}}      \\
{\tt F22287-1917}&      {\tt ESO 602-G025}&     {\tt 22h31m25.48s}&     {\tt -19d02m04.1s}&     {\tt 7507}&     {\tt 110.0}&    {\tt 11.34\phm{:}}      \\
{\tt F22389+3359}&      {\tt UGC 12150}&        {\tt 22h41m12.26s}&    {\tt +34d14m57.0s}&    {\tt 6413}&     {\tt 93.5}&     {\tt 11.35\phm{:}}      \\
{\tt F22467-4906}&      {\tt ESO 239-IG002}&    {\tt 22h49m39.87s}&     {\tt -48d50m58.1s}&     {\tt 12901}&    {\tt 191.0}&    {\tt 11.84\phm{:}}      \\
{\tt F22491-1808}& & {\tt 22h51m49.26s}&    {\tt -17d52m23.5s}&    {\tt 23312}&    {\tt 351.0}&    {\tt 12.20:}    \\
{\tt F23007+0836}&      {\tt NGC 7469}& {\tt 23h03m15.62s}&    {\tt +08d52m26.4s}&    {\tt 4892}&     {\tt 70.8}&     {\tt 11.65\phm{:}}      \\
{\tt F23024+1916}&      {\tt CGCG 453-062}&     {\tt 23h04m56.53s}&     {\tt +19d33m08.0s}&     {\tt 7524}&     {\tt 109.0}&    {\tt 11.38:}    \\
{\tt F23128-5919}&      {\tt ESO 148-IG002}&    {\tt 23h15m46.78s}&     {\tt -59d03m15.6s}&     {\tt 13371}&    {\tt 199.0}&    {\tt 12.06\phm{:}}      \\
{\tt F23135+2517}&      {\tt IC 5298}&  {\tt 23h16m00.70s}&     {\tt +25d33m24.1s}&     {\tt 8221}&     {\tt 119.0}&    {\tt 11.60\phm{:}}      \\
{\tt F23133-4251}&      {\tt NGC 7552}& {\tt 23h16m10.77s}&     {\tt -42d35m05.4s}&     {\tt 1608}&     {\tt 23.5}&     {\tt 11.11\phm{:}}      \\
{\tt F23157+0618}&      {\tt NGC 7591}& {\tt 23h18m16.28s}&     {\tt +06d35m08.9s}&     {\tt 4956}&     {\tt 71.4}&     {\tt 11.12\phm{:}}      \\
{\tt F23157-0441}&      {\tt NGC 7592}& {\tt 23h18m22.20s}&     {\tt -04d24m57.6s}&    {\tt 7328}&     {\tt 106.0}&    {\tt 11.40\phm{:}}      \\
{\tt F23180-6929}&      {\tt ESO 077-IG014}&    {\tt 23h21m04.53s}&     {\tt -69d12m54.2s}&     {\tt 12460}&    {\tt 186.0}&    {\tt 11.76\phm{:}}      \\
{\tt F23254+0830}&      {\tt NGC 7674}& {\tt 23h27m56.72s}&     {\tt +08d46m44.5s}&     {\tt 8671}&     {\tt 125.0}&    {\tt 11.56\phm{:}}      \\
{\tt 23262+0314 }&      {\tt NGC 7679}& {\tt 23h28m46.66s}&     {\tt +03d30m41.1s}&     {\tt 5138}&     {\tt 73.8}&     {\tt 11.11\phm{:}}      \\
{\tt F23365+3604}& & {\tt 23h39m01.27s}&     {\tt +36d21m08.7s}&     {\tt 19331}&    {\tt 287.0}&    {\tt 12.20:}    \\
{\tt F23394-0353}&      {\tt MCG -01-60-022}&   {\tt 23h42m00.85s}&     {\tt -03d36m54.6s}&     {\tt 6966}&     {\tt 100.0}&    {\tt 11.27\phm{:}}      \\
{\tt 23436+5257 }& &  {\tt 23h46m05.58s}&    {\tt +53d14m00.6s}&    {\tt 10233}&    {\tt 149.0}&    {\tt 11.57:}    \\
{\tt F23444+2911}&      {\tt Arp 86}&   {\tt 23h47m01.70s}&     {\tt +29d28m16.3s}&     {\tt 5120}&     {\tt 73.6}&     {\tt 11.07\phm{:}}      \\
{\tt F23488+1949}&      {\tt NGC 7771}& {\tt 23h51m24.88s}&     {\tt +20d06m42.6s}&     {\tt 4277}&     {\tt 61.2}&     {\tt 11.40\phm{:}}      \\
{\tt F23488+2018}&      {\tt MRK 0331}& {\tt 23h51m26.80s}&    {\tt +20d35m09.9s}&    {\tt 5541}&     {\tt 79.3}&     {\tt 11.50\phm{:}}      \\
\enddata
\tablenotetext{a}{When the IRAS Revised Bright Galaxy Sample (RBGS, Sanders et al. 2003)
was compiled, IRAS F10038-3338 was mistakenly cross-identified with the optical source
IC 2545. The proper optical counterpart is ESO 374-IG 032. (See the Essential Notes in NED.)}
%\tablenotetext{b}{When the RBGS was compiled, the best available heliocentric velocity for NGC 5010
%was $\rm 6400~km~s^{-1}$, adopted by NED from the RC3 (de Vaucouleurs et al. 1991).
%The correct value of $\rm V_{Helio} = 2975 (\pm~27)~km~s^{-1}$ (Wegner et al. 2003, via NED),
%drops the luminosity below the definition of a LIRG. The source is listed here because it was
%included in our {\it Spitzer} imaging and spectroscopic observations.}
\tablecomments{
Column (1): Original IRAS source, where an "F" prefix indicates the Faint Source Catalog and no prefix indicates the Point Source Catalog.
Column (2): Optical cross-identification, where available from NED. For many cases where the IRAS source corresponds to a pair of
optically identified galaxies, we adopt the system name instead of pair components. For example, IRAS F00163-1039 is identified in GOALS
as Arp 256 rather than "MCG -02-01-051/2" as in Sanders et al. 2003.
Column (3): The best available source right ascension (J2000) in NED as of October 2008.
Column (4): The best available source declination (J2000) in NED as of October 2008.
Column (5): The best available heliocentric redshift, expressed as a velocity, in NED as of October 2008.
Column (6): The luminosity distance in megaparsecs derived by correcting the heliocentric velocity for the 3-attractor flow model
of Mould et al. (2000) and adopting cosmological parameters
$\rm H_0 = 70~km~s^{-1}~Mpc^{-1}$, $\Omega_{Vacuum} = 0.72$, and $\Omega_{Matter} = 0.28$ based on the
five-year WMAP results (Hinshaw et al. 2009), as provided by NED.
Column (7): The total infrared luminosity in $\rm log_{10}$ Solar units computed using the flux densities reported
in the RBGS and the lumiosity distances in column (6) using the formulae
$L_{ir}/L_{\odot} = 4\pi (D_L [m])^2~(F_{ir}~[W~m^{-2}])/3.826\times 10^{26} [W m^{-2}]$,
where
$F_{ir} = 1.8\times 10^{-14}\{13.48 f_{12\mu m} + 5.16 f_{25\mu m} + 2.58 f_{60\mu m} + f_{100\mu m} [W m^{-2}]\}$
(Sanders \& Mirabel 1996); colons indicate uncertain IRAS measurements (Sanders, et al. 2003).
}
\end{deluxetable*}

\begin{deluxetable*}{lllc}
\label{tbl:data}
\tabletypesize{\footnotesize}
\setlength{\tabcolsep}{0.04in}
\tablenum{2}
\tablewidth{0pt}
\tablecaption{GOALS Data Sets}
\tablehead{
\colhead{\rm Type} & \colhead{Wavelength/Filter} & \colhead{Observatory} & \colhead{Number}\\
\colhead{(1)} & \colhead{(2)} & \colhead{(3)} & \colhead{(4)}
}
\startdata
X-Ray Imaging \& Spectroscopy& $0.4-7$ keV& Chandra - ACIS & 44\\
Ultraviolet imaging& 1528\AA (FUV), 2271\AA (NUV) & GALEX & 124\\
Ultraviolet imaging& 1400\AA (F140LP)& Hubble - ACS/SBC & 30\\
Ultraviolet imaging& 2180\AA (F218W)& Hubble - WFPC2& 30\\
Visual imaging& 4350\AA (F435W), 8140\AA (F814W)& Hubble - ACS & 88\\
Near-Infrared imaging& $1.6\mu$m (F160W) & Hubble - NICMOS & 88\\
Mid-Infrared imaging & 3.6, 4.5, 5.4, 8$\mu$m & Spitzer - IRAC & 202\\
Mid-Infrared nuclear spectroscopy & $5-40\mu$m low-res, $10-40\mu$m high-res & Spitzer - IRS & 202\\
Mid-Infrared spectral mapping & 5--40$\mu$m low-res & Spitzer - IRS & 42\\
Far-Infrared imaging  & 24,70,160$\mu$m & Spitzer - MIPS & 202\\
\enddata
\tablecomments{Summary of the primary (Spitzer, Chandra, Hubble and GALEX) GOALS data sets.  The type of data,
the wavelengths covered, the observatory, and the number of
LIRGs (systems) observed (including archival data) are given in cols. 1-4,
respectively.  Ancillary data is discussed in the text.}
\end{deluxetable*}

\end{document}